\documentclass[11pt,a4paper]{article}
\usepackage{jcappub}
\pdfoutput=1
\usepackage[english]{babel}
\usepackage{amsmath,amssymb,amsfonts, bm,bbm,slashed, subdepth}
\usepackage{graphicx}
\usepackage[sort&compress]{natbib}
\usepackage{xcolor}
\usepackage[normalem]{ulem}
\usepackage{hyperref}
\usepackage{cleveref}
\definecolor{red}{rgb}{1.0, 0, 0}
\usepackage{hyperref}
\usepackage{enumerate}
\usepackage{epsfig, subfigure}
\usepackage{setspace}
\usepackage{booktabs, tabularx}
\usepackage{units}
\usepackage{hyperref}

\newcommand{\be}{\begin{equation}}
\newcommand{\ee}{\end{equation}}
\newcommand{\ba}{\begin{array}}
\newcommand{\ea}{\end{array}}
\newcommand{\bea}{\begin{eqnarray}}
\newcommand{\eea}{\end{eqnarray}}
\newcommand{\balg}{\begin{align}}
\newcommand{\ealg}{\end{align}}
\newcommand{\bit}{\begin{itemize}}
\newcommand{\eit}{\end{itemize}}
\newcommand{\trm}[1]{\textrm{#1}}

\usepackage{enumitem}

\newcommand{\Mpc}{\trm{\Mpc}}
\newcommand{\yr}{\trm{\yr}}
\newcommand{\eV}{\trm{\eV}}

\begin{document}

\title{Nonlinear growth of structure in cosmologies with damped matter fluctuations}

\author[a]{Matteo Leo,}
\author[b]{Carlton M. Baugh,}
\author[b]{Baojiu Li}
\author[a]{and Silvia Pascoli}
\affiliation[a]{Institute for Particle Physics Phenomenology, Department of Physics, Durham University, South Road, Durham DH1 3LE, U.K.}
\affiliation[b]{Institute for Computational Cosmology, Department of Physics, Durham University, South Road, Durham DH1 3LE, U.K.}
\emailAdd{matteo.leo@durham.ac.uk}
\hfill{IPPP/17/98}

\abstract{We investigate the nonlinear evolution of structure in variants of the standard cosmological model which display damped density fluctuations relative to cold dark matter (e.g. in which cold dark matter is replaced by warm or interacting DM). Using N-body simulations, we address the question of how much information is retained from different scales in the initial linear power spectrum following the nonlinear growth of structure. We run a suite of N-body simulations with different initial linear matter power spectra to show that, once the system undergoes nonlinear evolution, the shape of the linear power spectrum at high wavenumbers does not affect the non-linear power spectrum, while it still matters for the halo mass function. Indeed, we find that linear power spectra which differ from one another only at wavenumbers larger than their half-mode wavenumber give rise to (almost) identical nonlinear power spectra at late times, regardless of the fact that they originate from different models with damped fluctuations. On the other hand, the halo mass function is more sensitive to the form of the linear power spectrum. Exploiting this result, we propose a two parameter model of the transfer function in generic damped scenarios, and show that this parametrisation works as well as the standard three parameter models for the scales on which the linear spectrum is relevant.} 
\maketitle

\section{Introduction}
Although the standard cold dark matter (CDM) paradigm successfully reproduces the observed structure of the Universe on large and intermediate scales, several studies have suggested tensions between the simplest CDM predictions and observations on small scales (for a recent review of the small scale problems in CDM see \cite{Weinberg:2013aya}). CDM haloes display central density cusps in N-body simulations \cite{Dubinski:1991bm,Navarro:1995iw,Navarro:1996gj}. Such profiles appear to be at odds with the observed small scale dynamics of some spiral galaxies, which suggest a constant DM distribution (core) in the center \cite{deBlok:2001hbg, Salucci:2011ee}. Moreover, a large difference was found between the number of satellite galaxies observed in the Milky Way and the number of subhaloes in CDM simulations \cite{Klypin:1999uc,Moore:1999nt} (the so-called ``missing satellites'' problem). This observed lack of small structures implies that galaxy formation only takes place in the most massive MW subhaloes, but when we look at these structures they appear to be less dense than expected in CDM simulations (the so-called ``too big to fail'' problem, first identified by \cite{2011MNRAS.415L..40B}). Several solutions have been proposed to ameliorate these shortcomings within the CDM paradigm, e.g. by taking into account baryonic physics \cite{Mashchenko:2007jp,2012MNRAS.421.3464P,2013MNRAS.432.1947M,2014ApJ...786...87B}. The absence of massive subhaloes in the Milky Way could be also interpreted as an indication that the MW halo is less massive than is commonly assumed \cite{2012MNRAS.424.2715W}. 

Due to these difficulties alternatives to the standard $\Lambda$CDM have been proposed which display less power on small scales; we will refer to these as {\it damped models}. The mechanisms leading to a suppression of the power on small scales can be divided in two broad classes, those that involve modifications of the primordial power spectrum on small scales in the early Universe (achieved by modifications of the inflaton dynamics) \cite{Kamionkowski:1999vp,White:2000sy,Yokoyama:2000tz,Zentner:2002xt,Ashoorioon:2006wc,Kobayashi:2010pz,Nakama:2017ohe} and those that suppress the power at later times (due to some non-standard DM mechanism) \cite{Bode:2000gq,Colin:2000dn, Viel:2005qj,Hansen:2001zv,Dodelson:1993je,Dolgov:2000ew,Asaka:2006nq,Enqvist:1990ek,Shi:1998km,Abazajian:2001nj,Kusenko:2006rh,Petraki:2007gq,Merle:2015oja,Konig:2016dzg,Boehm:2004th,Boehm:2014vja,Schewtschenko:2014fca,Spergel:1999mh,Marsh:2015xka}. The standard $\Lambda$CDM scenario has a nearly scale invariant primordial power spectrum and cold (non-interacting and massive) dark matter particles, so matter fluctuations are present on all scales. To ensure damping in the matter power spectrum,  the models of the first class are characterised by  a breaking of the scale invariance of the primordial spectrum on subgalactic scales (e.g. in a single-field scenario, the damping is obtained by a suitable choice of the inflaton potential \cite{Kamionkowski:1999vp}), while the standard cold dark matter sector remains unchanged (in these models the DM particles are still massive and non-interacting). The suppression of the power in these models is achieved before the radiation-dominated era.  The models in the second class, on the other hand, introduce non-standard DM mechanisms (that modify the shape of the power spectrum during the evolution of the fluctuations), while the primordial power spectrum is scale-invariant. These models are commonly dubbed ``non-cold'' dark matter (hereafter nCDM) scenarios. The mechanism leading to a suppression of power in nCDM depends on the particular particle production process. Nevertheless, nCDM candidates are often characterised either by a non-negligible thermodynamic velocity dispersion (the so-called warm DM models  \cite{Bode:2000gq,Colin:2000dn, Viel:2005qj,Hansen:2001zv,Dodelson:1993je,Dolgov:2000ew,Asaka:2006nq,Enqvist:1990ek,Shi:1998km,Abazajian:2001nj,Kusenko:2006rh,Petraki:2007gq,Merle:2015oja,Konig:2016dzg}), interactions (DM interacting with standard model particles such as neutrinos or photons \cite{Boehm:2004th,Boehm:2014vja,Schewtschenko:2014fca} and self-interacting DM \cite{Spergel:1999mh}) or pressure terms from macroscopic wave-like behaviour (e.g. ultra-light axions \cite{Marsh:2015xka}). 

All of the phenomena described above introduce a characteristic scale below which the density fluctuations are erased resulting in a cut-off appearing in the linear matter power spectrum. Because of this difference with respect to the standard cosmological model, it is important to investigate how the predictions for structure formation differ in damped models from those in the standard $\Lambda$CDM. N-body simulations have proved to be a powerful tool to model the nonlinear evolution of cosmic structure in the standard $\Lambda$CDM scenario and can also be used to study the effects of the damping on small scales. However, different damped models display different forms for the linear power spectrum, whose shape and cut-off position depend on the particular model. This implies that, in principle, one needs to analyse the entire plethora of power spectra (each coming from a specific scenario and a specific set of particle physics parameters) to study the impact of every single damped model on structure formation. In \cite{Murgia:2017lwo}, the authors found a general parametrisation of the power spectrum with three free parameters, which is flexible enough to reproduce accurately the linear power spectra of a large class of nCDM models\footnote{We note that although this formula was not directly tested on models which belong to the first class, it should work for these since some of the models with primordial broken scale invariance produce linear power spectra similar to nCDM models (e.g. the step-type suppression in the primordial power spectrum considered in \cite{Nakama:2017ohe}  generates a linear matter power spectrum similar to that of a mixed DM model).}.

However, the nonlinear evolution of structure transfers power from large scales to small scales, so the differences at small scales between standard $\Lambda$CDM and the damped models can be significantly reduced in the nonlinear matter power spectra (see e.g. \cite{2012MNRAS.421...50V,Leo:2017zff}). It is therefore interesting to understand how well different damped models can be distinguished from the point of view of cosmological structure measurements and if there is a limit to how much simulations can tell us about different models. In other words, if two damped linear spectra are very similar to one another on large scales and they only differ appreciably at small scales (say, below the half-mode wavelength, see the next section for a definiton of this quantity), we can ask if such differences survive the nonlinear growth of structure and what imprint, if any, they leave behind. If gravitational instability erases these differences, the two spectra (although coming from two different theories) give the same results in terms of cosmological observables at late times and, as a consequence, this limits what we can learn about the nature of dark matter from large-scale structure\footnote{Indeed, large scale structure cannot distinguish if the cut-off in the matter power spectrum is due to a primordial damping or a late time DM mechanism. Other observables could be more sensitive to the physics of the early Universe \cite{Nakama:2017ohe}.}.

Here, we show to what extent the full shape of a linear damped power spectrum influences structure formation. We run a series of N-body simulations starting from different initial linear matter power spectra which are identical on large scales but differ substantially on small scales. One of our aims is to establish down to which scale does the shape of the linear power spectrum matter as regards the nonlinear growth of structure. We measure the nonlinear matter power spectra from simulations at low redshifts and compare between one another and with standard CDM results. Moreover, we analyse halo catalogues extracted from the simulations, with particular attention to how the halo mass function is influenced by the initial power spectrum. Based on our results, we propose a 2-parameter model for the initial power spectrum (which we compare against the 3-parameter model found in \cite{Murgia:2017lwo}) and show that two parameters are sufficient to capture the interesting features of the power spectrum from the point of view of structure formation. 

The paper is organised as follows. In Sections 2 and 3, we describe the initial power spectra used in our analysis and the set-up of the N-body simulations. In Section 4, we present the results from our simulations, measuring nonlinear power spectra and the halo mass function for several damped models. Finally, the description and the results from our 2-parameter fitting formula are presented in Section 5. We conclude in Section 6. A companion paper discusses improvements to the analytic calculation of the halo mass function to match simulation results including those presented in this paper \cite{Leo:2018odn}.

\section{Initial power spectra}\label{sec:IPS}

\begin{figure}
\centering
{\includegraphics[width=.8\textwidth]{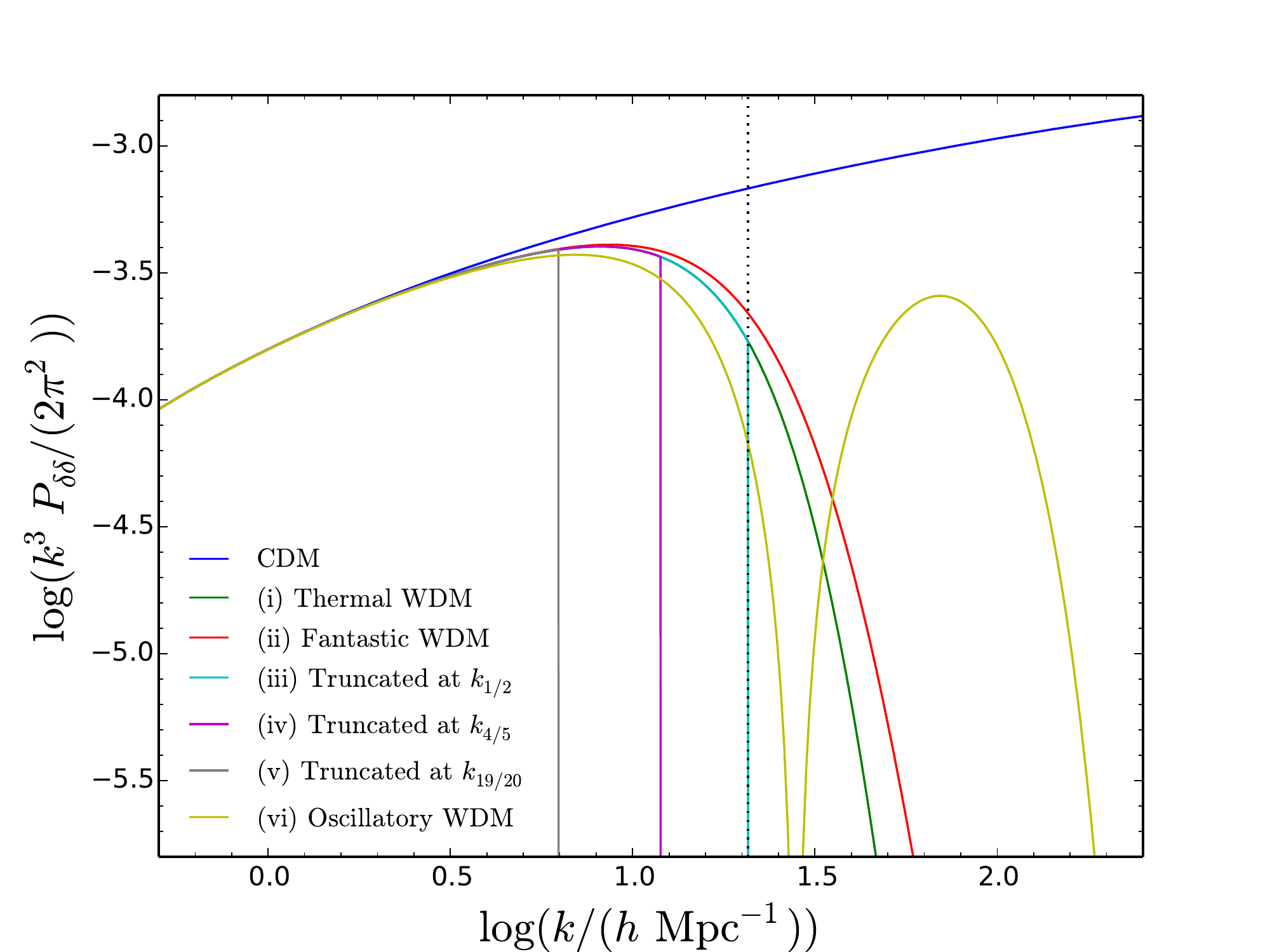}}
\caption{Initial linear perturbation theory matter power spectra at $z=199$ for different models as labelled. The black vertical dotted line represents the half-mode wavenumber $k_{1/2}$ for the thermal WDM power spectrum (green). The power spectra are described in Section \ref{sec:IPS}.}
\label{fig:alllinearmatterspectra}
\end{figure}

In this section we present the linear power spectra used to generate the initial conditions (ICs) for the N-body simulations. We consider the following initial damped power spectra (some of these are phenomenological, i.e. specific linear power spectra which have no theoretical motivation, but are considered here as test cases). 
\begin{enumerate}[label=(\roman*)]
\item {\it Thermal WDM} -- The matter power spectrum from a thermal warm DM (WDM) candidate. For this model the power spectrum can be well approximated\footnote{More accurate power spectra for more general non-cold DM models can be generated using Boltzmann codes such as {\sc class} \cite{2011arXiv1104.2932L,2011JCAP...09..032L}.} by the parametrisation in \cite{Bode:2000gq}. This parametrisation can be written in terms of a transfer function relative to standard CDM, $T^2(k) \equiv P^{\mathrm{WDM}}(k)/P^{\mathrm{CDM}}(k)$:
\begin{equation}
T(k) = \left(1 + \left(\alpha k\right)^{\beta}\right)^{\gamma},
\label{eq:fittingformula}
\end{equation}
where 

\begin{equation}
\alpha = a  \left(\frac{\Omega^0_\mathrm{WDM}}{0.25}\right)^b  \left(\frac{h}{0.7}\right)^c  \left(\frac{m_\mathrm{WDM}}{\text{keV}}\right)^d, \quad \beta=2 \nu, \quad \gamma=-5 / \nu,
\label{eq:parametersthermal}
\end{equation}
and 
\[
a = 0.049, \quad b = 0.11, \quad c = 1.22, \quad d = - 1.11, \quad \nu = 1.12,
\]
as presented in \cite{Viel:2005qj}. The WDM candidate is chosen to have mass $m_\mathrm{WDM} =2$ keV. We introduce three characteristic scales: $k_{1/2},k_{4/5}$ and $k_{19/20}$. $k_{1/2}$ is the half-mode wavenumber at which the transfer function (eq. (\ref{eq:fittingformula})) is suppressed by 50\%, i.e. $T =1/2$. While $k_{4/5}$ and $k_{19/20}$ are the wavenumbers at which $T = 4/5 $  and $T = 19/20$ respectively, i.e. at these wavenumbers the transfer function is suppressed by $20\%$ and $5\%$ with respect to CDM. Given the parametrisation in eq. (\ref{eq:fittingformula}) for $T(k)$ and for a mass of $m_\mathrm{WDM}=2$ keV, we have 
\begin{equation}
k_{1/2} \simeq 21.2 \,h\, \mathrm{Mpc}^{-1},
\label{eq:halfmodescale}
\end{equation} 
\begin{equation}
k_{4/5} \simeq 12.2 \,h\, \mathrm{Mpc}^{-1},
\label{eq:4/5scale}
\end{equation} 
\begin{equation}
k_{19/20} \simeq 6.4 \,h\, \mathrm{Mpc}^{-1},
\label{eq:19/20scale}
\end{equation} 
so, $k_{1/2}>k_{4/5}>k_{19/20}$.
\item {\it Fantastic WDM} -- We generate a second power spectrum from the parametrisation given in eq. (\ref{eq:fittingformula}), by fixing the parameters at $\alpha= 0.0233$, $\beta=2.128$ and $\gamma =-2.946$. We choose this power spectrum because it is identical to the power spectrum for thermal WDM in case (i) at small wavenumbers, while it starts to differ at scales beyond $k\sim k_{4/5}$ (the differences in shape of these two power spectra are very similar to those between a resonantly-produced sterile neutrino and a thermal WDM with candidate mass $m_\mathrm{WDM}=3.3$ keV, e.g. see figure 1 in \cite{Bose:2015mga}).
\item {\it Truncated at $k_{1/2}$} -- The third initial linear power spectrum is obtained by truncating the power spectrum in (i) at $k=k_{1/2}$ such that for $k\leq k_{1/2}$ the $P(k)$ for cases (i) and (iii) are identical, while for $k>k_{1/2}$ the (iii) power spectrum is $P(k>k_{1/2})=0$, 
\begin{equation}
P_{\mathrm{(iii)}}(k)=\begin{cases} P_{\mathrm{(i)}}(k) \quad &\mathrm{if} \quad k\leq k_{1/2}\\
0 \quad &\mathrm{if} \quad k>k_{1/2}.\\
\end{cases}
\end{equation} 
\item {\it Truncated at $k_{4/5}$} -- The fourth power spectrum is obtained by truncating the power spectrum in (i) at $k=k_{4/5}$ such that for $k\leq k_{4/5}$ the $P(k)$ for cases (i) and (iv) are identical, while for $k>k_{4/5}$ the (iv) power spectrum is $P(k>k_{4/5})=0$,
\begin{equation}
P_{\mathrm{(iv)}}(k)=\begin{cases} P_{\mathrm{(i)}}(k) \quad &\mathrm{if} \quad k\leq k_{4/5}\\
0 \quad &\mathrm{if} \quad k>k_{4/5}.\\
\end{cases}
\end{equation} 
\item {\it Truncated at $k_{19/20}$} -- The fifth power spectrum is obtained by truncating the power spectrum in (i) at $k=k_{19/20}$ such that for $k\leq k_{19/20}$ the $P(k)$ for cases (i) and (v) are identical, while for $k>k_{19/20}$ the (v) power spectrum is $P(k>k_{19/20})=0$,
\begin{equation}
P_{\mathrm{(v)}}(k)=\begin{cases} P_{\mathrm{(i)}}(k) \quad &\mathrm{if} \quad k\leq k_{19/20}\\
0 \quad &\mathrm{if} \quad k>k_{19/20}.\\
\end{cases}
\end{equation}
\item {\it Oscillatory WDM} -- Our last power spectrum is an oscillatory one. This spectrum is inspired by interacting DM \cite{Boehm:2004th,Boehm:2014vja,Schewtschenko:2014fca}, but we have artificially enhanced the amplitude of the first peak to see if there are any signatures of the oscillation after the nonlinear growth of structure.
\end{enumerate}
All the linear power spectra are shown in Figure~\ref{fig:alllinearmatterspectra}, together with that for standard CDM. Here, we present the matter power spectra at $z=199$ plotted as $\Delta(k) \equiv k^3 P(k)/(2\pi^2)$. We stress that the power spectra in cases (ii-vi) are not physically motivated but instead are intended to test how changing the shape of the initial linear power spectrum influences the nonlinear evolution of structure.

\section{The simulations}
The linear matter power spectra in Figure~\ref{fig:alllinearmatterspectra} are used to generate the initial conditions (ICs) for N-body simulations. We use the 2LPTic code \cite{Crocce:2006ve}, which provides ICs based on second-order Lagrangian perturbation theory. The initial redshift is chosen to be $z_\mathrm{ini}=199$, at which all the wavenumbers probed in the simulation are well inside the linear regime. The simulations are performed in a cubic box of comoving length $L=25$ $h^{-1}$Mpc using $N=512^3$ particles. We choose this pair of $\{N,L\}$ in our simulations since we want to resolve the structures on scales near the half-mode wavenumber of the power spectrum for a thermal WDM candidate with mass $m_\mathrm{WDM} =2 $ keV (see Figure~\ref{fig:alllinearmatterspectra}). The Nyquist frequency of a simulation is $k_{Ny} \equiv \pi (N^{1/3}/L)$ (this specifies the value up to which we can trust the $P(k)$). We evolve the ICs to $z=0$, using the publicly available tree-PM code Gadget-2 \cite{Springel:2005mi}. The gravitational softening length $\epsilon$ is set to be $1/40$-th of the mean interparticle separation, $L/N^{1/3}$. We also note that in our simulations we do not include thermal velocities because their physical effects are negligible for our choice of WDM candidate masses and N-body parameters (see e.g. \cite{Schneider:2013ria,2012MNRAS.420.2318L,2013MNRAS.428..882M,2013MNRAS.430.2346S}), and including them introduces extra numerical noise in the simulations, reducing the range of scales we can trust \cite{Leo:2017zff}.

\section{Results from N-body simulations}

\begin{figure}[h]
\centering
\subfigure[][]
{\includegraphics[width=.75\textwidth]{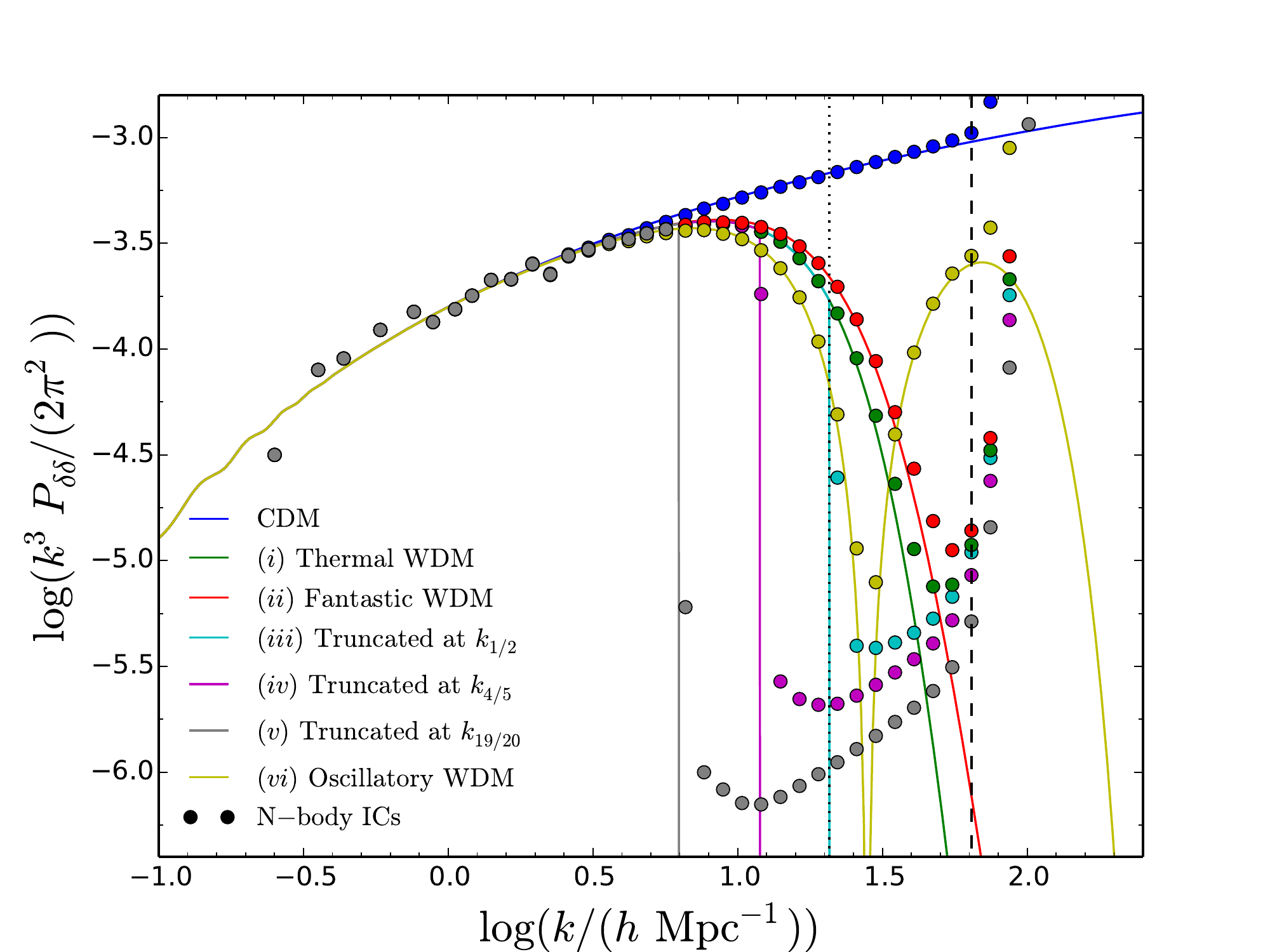}\label{fig:alllinearmatterspectraNbodya}}
\subfigure[][]
{\includegraphics[width=.75\textwidth]{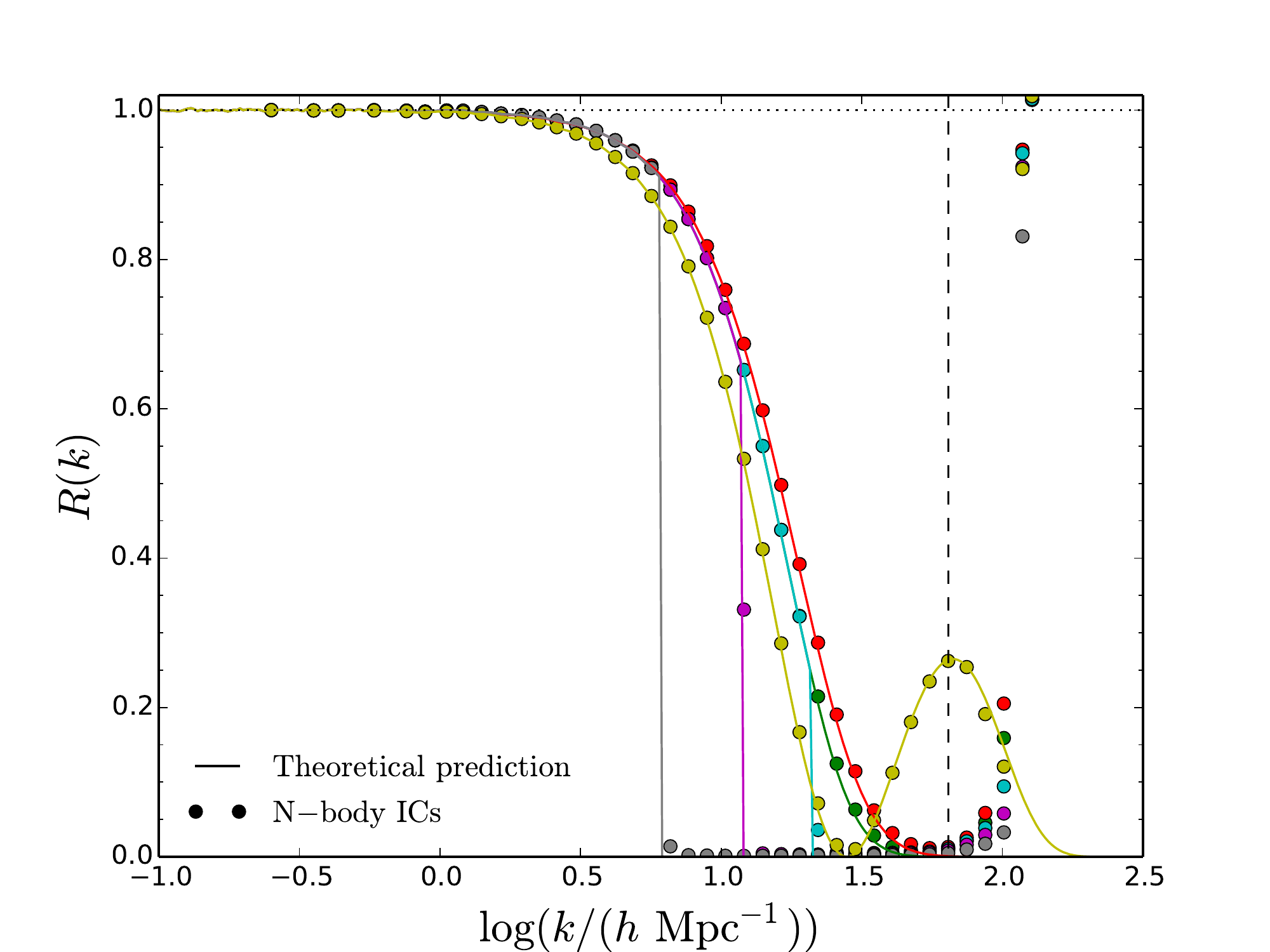}\label{fig:alllinearmatterspectraNbodyb}}
\caption{(a) Initial linear matter power spectra generated at $z=199$. The symbols represent the matter power spectra measured from the ICs. The black vertical dotted line represents the half-mode wavenumber $k_{1/2}$ for the thermal WDM power spectrum (green curve). (b) Ratios of damped power spectra at $z=199$ relative to that from CDM, see Eq.~(\ref{eq:relativeratio}). The black dashed line in both panels indicates the Nyquist frequency of the simulations. The colour scheme is the same as used in Fig.~1.}
\label{fig:alllinearmatterspectraNbody}

\end{figure}

\subsection{Matter power spectra}
We measure the matter power spectrum using a code based on the cloud-in-cell mass assignment scheme. The matter power spectra measured from the ICs are shown in Figure~\ref{fig:alllinearmatterspectraNbodya} as symbols, while the lines show the theoretical power spectra. In Figure~\ref{fig:alllinearmatterspectraNbodyb} we show the ratio,
\begin{equation}
R(k) = \frac{P_\mathrm{damped}(k)}{P_\mathrm{CDM}(k)},
\label{eq:relativeratio}
\end{equation}
where $P_\mathrm{damped}(k)$ is the matter power spectrum measured from a simulation of a particular damped model, while $P_\mathrm{CDM}(k)$ is that measured from the $\Lambda$CDM simulation. As we can see, the measured matter power spectra are in good agreement with the theoretical expectations up to the Nyquist frequency. We also note that although the power spectra are very similar at small wavenumbers, they differ appreciably from one another at high wavenumbers, e.g. the ratio between thermal and fantastic WDM at $k\sim k_{1/2}$ is $\sim 0.75$ and reaches $\sim 6$ between thermal WDM and the truncated at $k_{1/2}$ power spectra. For the cases of spectra truncated at $k_{4/5}$ and $k_{19/20}$ the deviations start at even smaller wavenumbers.

\begin{figure}
\centering
\subfigure[][$z=9$]
{\includegraphics[width=.495\textwidth]{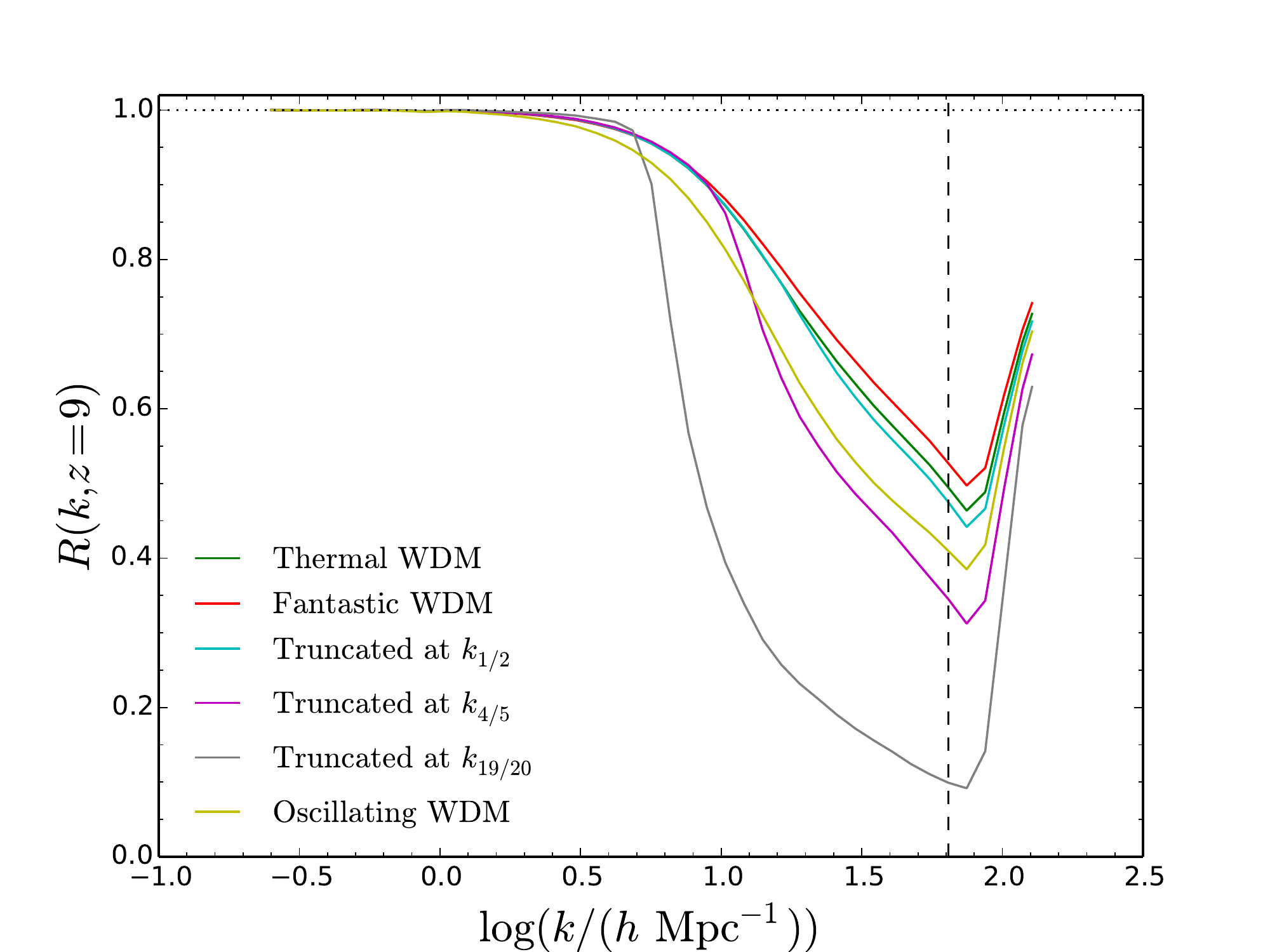}}
\subfigure[][$z=5$]
{\includegraphics[width=.495\textwidth]{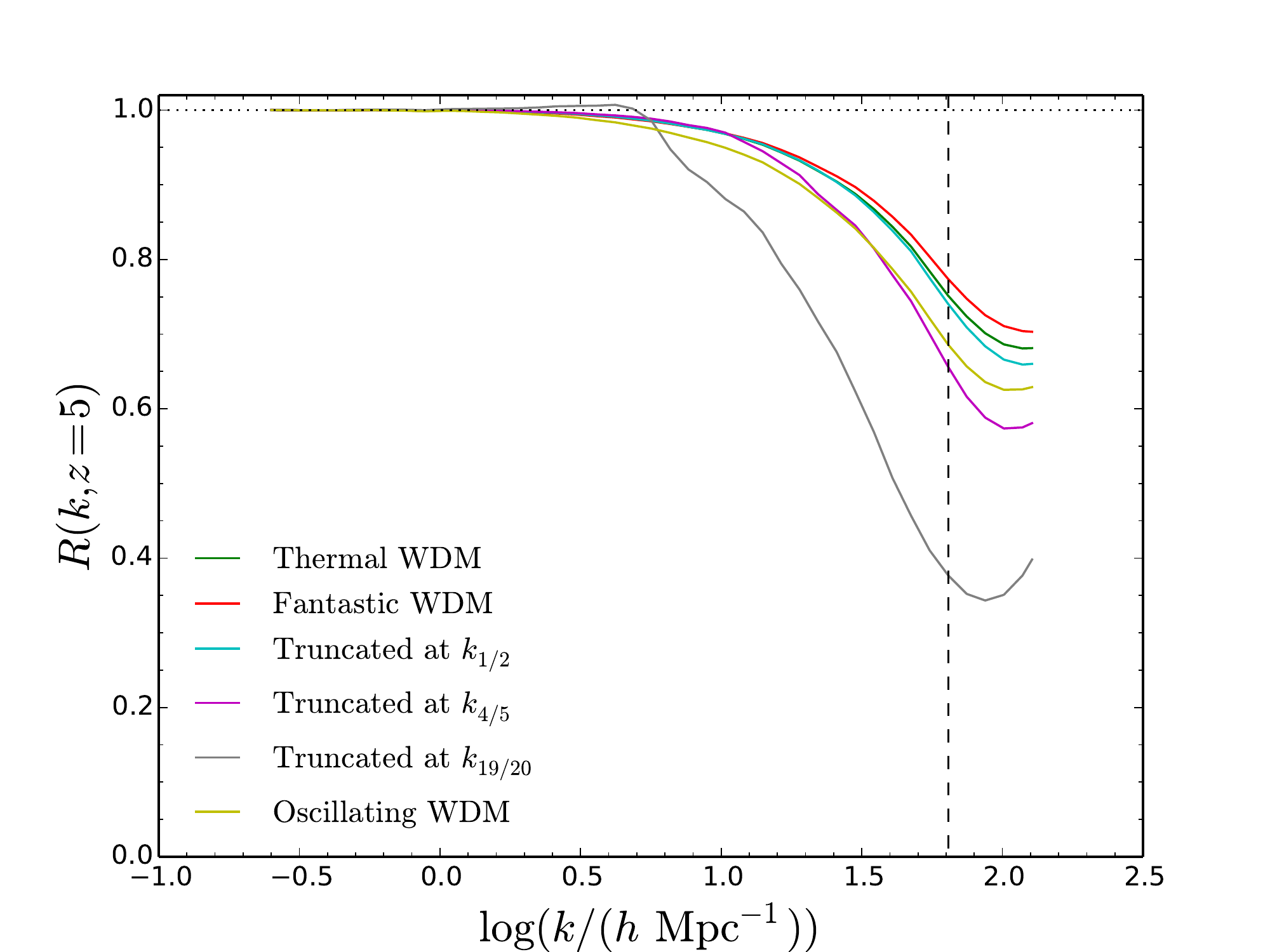}}\\
\subfigure[][$z=3$]
{\includegraphics[width=.495\textwidth]{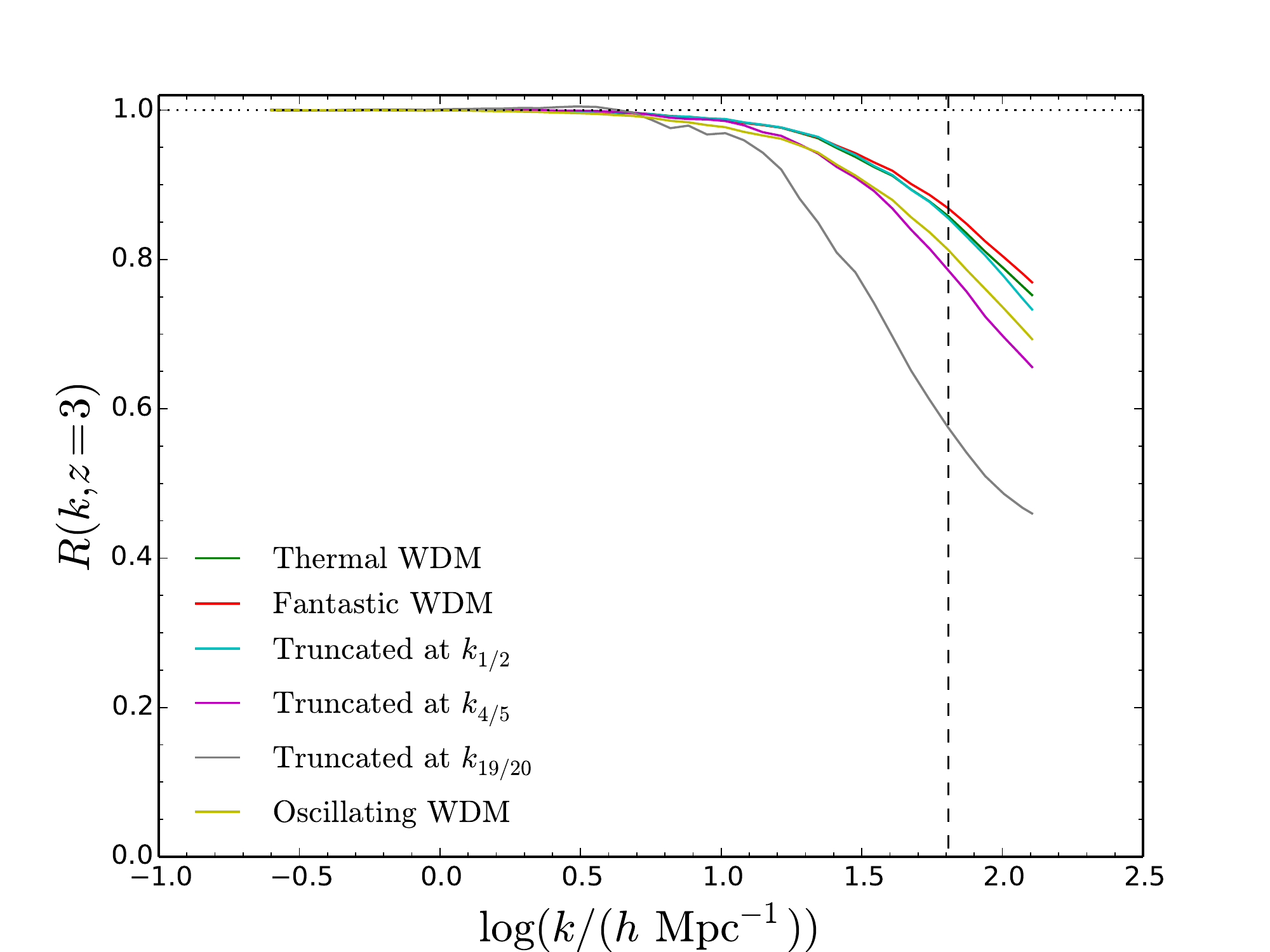}}
\subfigure[][$z=1$]
{\includegraphics[width=.495\textwidth]{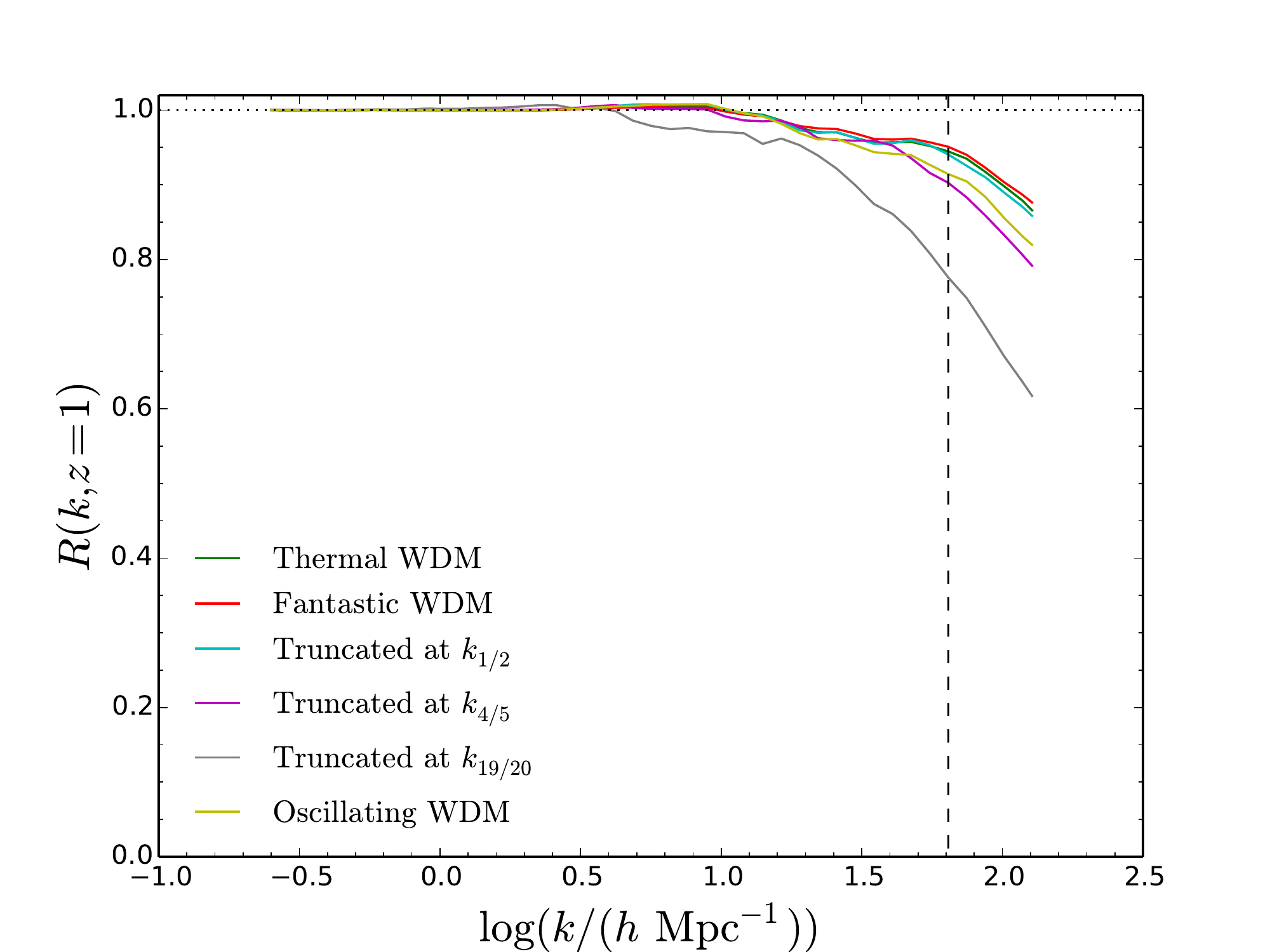}}
\subfigure[][$z=0$]
{\includegraphics[width=.495\textwidth]{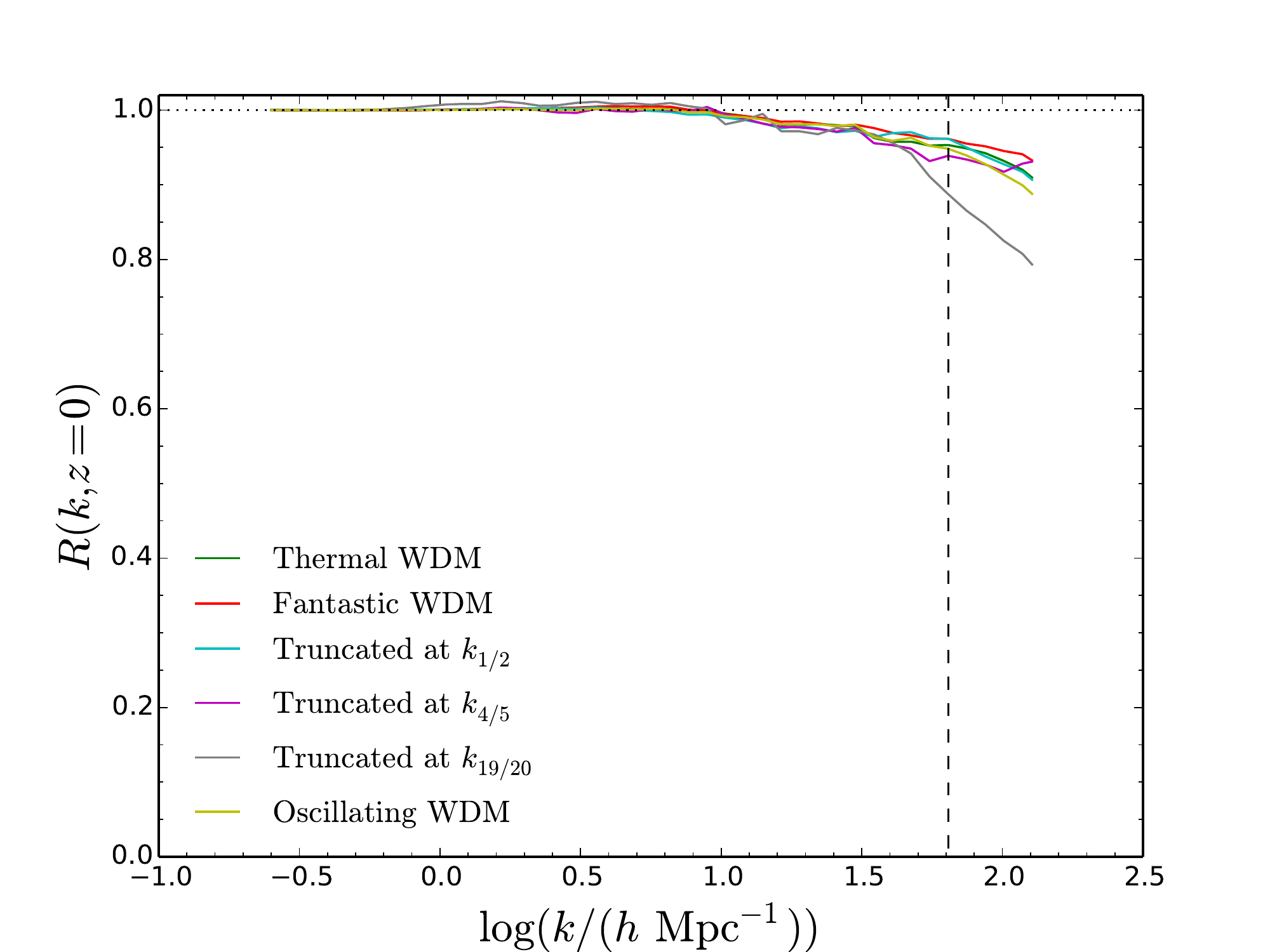}}
\caption{Ratios of matter power spectra measured from simulations with damped fluctuations with respect to those measured from a CDM simulation. Each panel shows a different redshift as labelled. The black vertical dashed line displays the Nyquist frequency of the simulations.}
\label{fig:alllinearmatterspectraratios}

\end{figure}

The situation changes when the ICs are evolved and the system undergoes nonlinear evolution. In Figure~\ref{fig:alllinearmatterspectraratios} we plot the matter power spectra ratios (see eq. (\ref{eq:relativeratio})) for (evolved) simulations at redshifts $z=9,5,3,1,0$. We can see that the transfer of power from large to small scales progressively reduces the differences between CDM and the various damped scenarios and the difference between the $P(k)$ in the different damped models themselves decreases in time. Indeed, at $z=0$ for all the spectra, except for the most extreme truncation at $k_{19/20}$, the relative difference between each damped spectrum respect to the thermal WDM one is always less than $\sim 1.2\%$  over all the scales resolved by the simulations. We find also that the differences between the power spectrum for the model truncated at $k_{19/20}$ and the other damped power spectra become progressively smaller at lower redshifts (they never exceed $\sim 8\%$ at $z=0$) and the differences are pushed to higher wavenumbers. 

The oscillatory pattern in the linear power spectrum of the oscillatory WDM, the first peak of which is well resolved in the N-body ICs (see Figure~\ref{fig:alllinearmatterspectraNbody}), is washed away during the nonlinear evolution; we can see in Figure~\ref{fig:alllinearmatterspectraratios} that no trace of it remains at late times. This suggests that the nonlinear power spectrum cannot be used to distinguish between models with damped fluctuation spectra, such as warm and interacting dark matter.  

\subsection{Halo mass function}
The story is different when we look at the halo mass function. This quantity is more sensitive to the initial conditions and the form of the linear theory power spectrum than the evolved power spectrum at late times and, indeed, we find appreciable differences in  the halo catalogues extracted from N-body simulations of different damped models. To extract the halo properties, we use the code {\sc rockstar}, which is a phase-space friends-of-friends halo finder \cite{2013ApJ...762..109B}. As a definition of the halo mass, we use the mass, $M_{200}$, contained in a sphere of radius $r_{200}$, within which the average density is $200$ times the critical density of the universe at the specified redshift. The (differential) halo mass function is presented as $F(M_\mathrm{200}, z) = {\rm d}n/{\rm d}\log(M_\mathrm{200})$, where $n$ is the number density of haloes with mass $M_\mathrm{200}$. 

\begin{figure}
\centering
{\includegraphics[width=.83\textwidth]{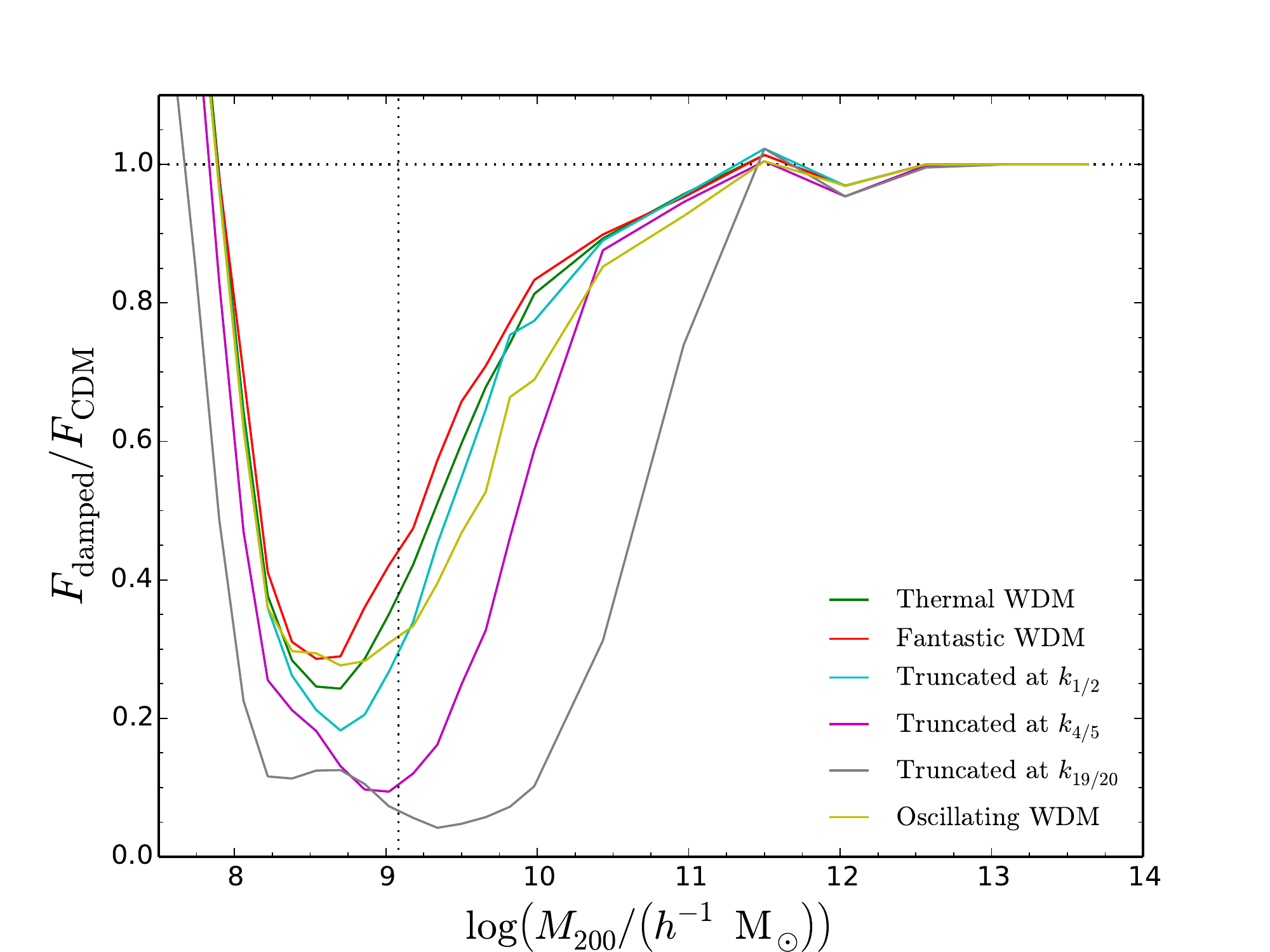}}
\caption{Ratios of the halo mass function measured from N-body simulations at $z=0$ for the various damped models (as labelled) respect to that of CDM. The black vertical dotted line represents the half-mode mass for the thermal WDM model, $M_\mathrm{hm}= ({4\pi}/{3}) \,\bar{\rho} \left({\pi}/{k_{1/2}}\right)^3$, where $\bar{\rho}$ is the mean density of the universe and $k_{1/2}$ is the half-mode wavenumber of the thermal WDM $P(k)$.}
\label{fig:alllinearmatterspectrahaloes}

\end{figure}

In Figure~\ref{fig:alllinearmatterspectrahaloes} we show the ratios between the damped and CDM halo mass function at $z=0$ extracted from the simulations. As we can see, the six damped spectra give rise to halo mass functions which are noticeably different from one another and from CDM. For example, the ratio between the measured halo mass function from a thermal WDM model and scenario in which the initial power spectrum is truncated at $k_{19/20}$ is around a factor of $8$ at a halo mass of $M_{200}\sim 10^{10} \,h^{-1}\,\mathrm{M}_\odot$. This is remarkable as the ratio in the nonlinear matter power spectra between the two models at $z=0$ never exceeds $\sim 1.08$ (see previous subsection). Similar conclusions are reached for the other damped models, although for them the differences in the halo mass function with respect to the thermal WDM case are less pronounced.

We note that for $M_{200}<10^9\,h^{-1}\, \mathrm{M}_\odot$, the halo catalogues measured from our N-body simulations of damped models are dominated by spurious haloes (visible as an upturn in the halo mass function, see Figure~\ref{fig:alllinearmatterspectrahaloes}). Spurious haloes are numerical artefacts that, in general, appear in simulations of damped models in which the initial power spectrum has a resolved cut-off, see e.g. \cite{Bode:2000gq,Wang:2007he,Lovell:2013ola,Schewtschenko:2014fca,2012MNRAS.424..684S,Power:2013rpw,Power:2016usj,Schneider:2013ria,Bose:2015mga,Schneider:2014rda}. These unphysical structures can be identified (and then removed), e.g. using the method described in \cite{Lovell:2013ola}. Since interested in the overall behaviour of the halo mass function for $M_{200}>10^9\,h^{-1}\, \mathrm{M}_\odot$, here we will not go into a detailed study of how to eliminate spurious haloes. We refer to a companion study \cite{Leo:2018odn}, where we have explored these features in more detail and where we have shown results from cleaned halo catalogues for the models studied here.

\section{2-parameter transfer function}
In \cite{Murgia:2017lwo}, the authors proposed an analytical parametrisation for the power spectrum which is flexible enough to match the linear theory matter power spectra for a wide range of nCDM models discussed in the literature (e.g. non-resonantly and resonantly produced sterile neutrinos, sterile neutrinos from scalar decay, ultra-light axions, etc). The mathematical form of this parametrization is identical to that used for thermal WDM (see Eq.~(\ref{eq:fittingformula})). However, unlike the thermal WDM case, the three parameters $\{\alpha,\beta,\gamma\}$ are not related to one another and are left free (hereafter we call the parametrization introduced by \cite{Murgia:2017lwo} the 3-parameter transfer function). 

In the previous section we saw that after the nonlinear evolution of structure the shape of the linear power spectrum for $k>k_{1/2}$ becomes unimportant in determining the nonlinear power spectrum at intermediate and low redshifts. Moreover, if two linear spectra differ minimally at $k\gtrsim k_{1/2}$, we find no appreciable deviation in the halo mass function predicted by the models (as will be confirmed below). So, since the full shape of the linear matter power spectrum is irrelevant from the point of view of structure formation\footnote{We agree that our results are strictly true only for linear theory $P(k)$ with $k_{1/2}$ around or larger than the values considered here. This means that our results may not apply in the case of linear $P(k)$ with half-mode wavenumbers smaller than those considered here. However, we note that e.g. a thermal WDM matter candidate with $m_\mathrm{WDM}<2$ keV is strongly disfavoured by the current Lyman-$\alpha$ constraints \cite{Viel:2013apy,Irsic:2017ixq}. So, our results can be considered to be general in the sense that they can be applied to all the damped models which are not already ruled out by astrophysical constraints.}, we can ask if the number of free parameters in the parametrisation found in \cite{Murgia:2017lwo} can be reduced if we are interested only in the form of the power spectrum for $k\leq k_{1/2}$. Indeed, out of the three parameters $\gamma$ is the one which controls the slope of $T(k)$ for $k>k_{1/2}$ \cite{Murgia:2017lwo}, so it seems reasonable to reduce the number of parameters by fixing the value of $\gamma$. Here, we fix $\gamma$ such that it is equal to the value in the case of thermal WDM, i.e. $\gamma = -5/\nu$ with $\nu = 1.12$ (see the value of $\gamma$ in (\ref{eq:parametersthermal})). Our parametrisation will then read
\begin{equation}
T(k) = \left(1 + \left(\tilde{\alpha} k\right)^{\tilde{\beta}}\right)^{-5/\nu},
\label{eq:fittingformularevisited}
\end{equation}
where $\{\tilde{\alpha},\tilde{\beta}\}$ are the new free parameters. This new parametrisation has only two free parameters (hereafter we call it the 2-parameter transfer function). The two new parameters $\{\tilde{\alpha},\tilde{\beta}\}$ are in general different from the old ones $\{{\alpha},{\beta}\}$. This is because although $\gamma$ is mostly responsible for the shape of the transfer function at $k>k_{1/2}$, it also makes some contribution to $T(k)$ at $k\leq k_{1/2}$. So, in order to capture the slope of the 3-parameter transfer function at small wavenumbers, the free parameters in the new parametrisation need to be different from the old $\{{\alpha},{\beta}\}$. We show below that this new parametrisation is able to match very well the 3-parameter fitting function for $k\leq k_{1/2}$.

\begin{table}[t]
\centering
\begin{tabular}{c|c|c}
\toprule
Model &3-parameter transfer function & 2-parameter transfer function \\
&$\alpha$, $\beta$ , $\gamma$&$\tilde{\alpha}$, $\tilde{\beta}$\\
\midrule
Resonantly Produced (I)& $0.025$, $2.3$, $-2.6$ & $0.019$, $2.250$\\
Resonantly Produced (II)& $0.071$, $2.3$, $-0.98$ & $0.029$, $2.029$\\
Scalar Decay & $0.016$, $2.6$, $-8.1$ & $0.021$, $2.637$\\
Non-resonantly produced & $0.038$, $2.2$, $-4.4$ & $0.037$, $2.199$\\
\bottomrule
\end{tabular}
\caption{Values of the three parameters $\{\alpha,\beta,\gamma\}$ found in \cite{Murgia:2017lwo} and of our two parameters $\{\tilde{\alpha},\tilde{\beta}\}$ for the transfer function of the models listed in the first column.}
\label{tab:param}
\end{table}

\begin{figure}[]
\advance\leftskip-1.cm
\advance\rightskip-2cm
\subfigure[][Resonantly produced (I), transfer functions.]
{\includegraphics[width=.6\textwidth]{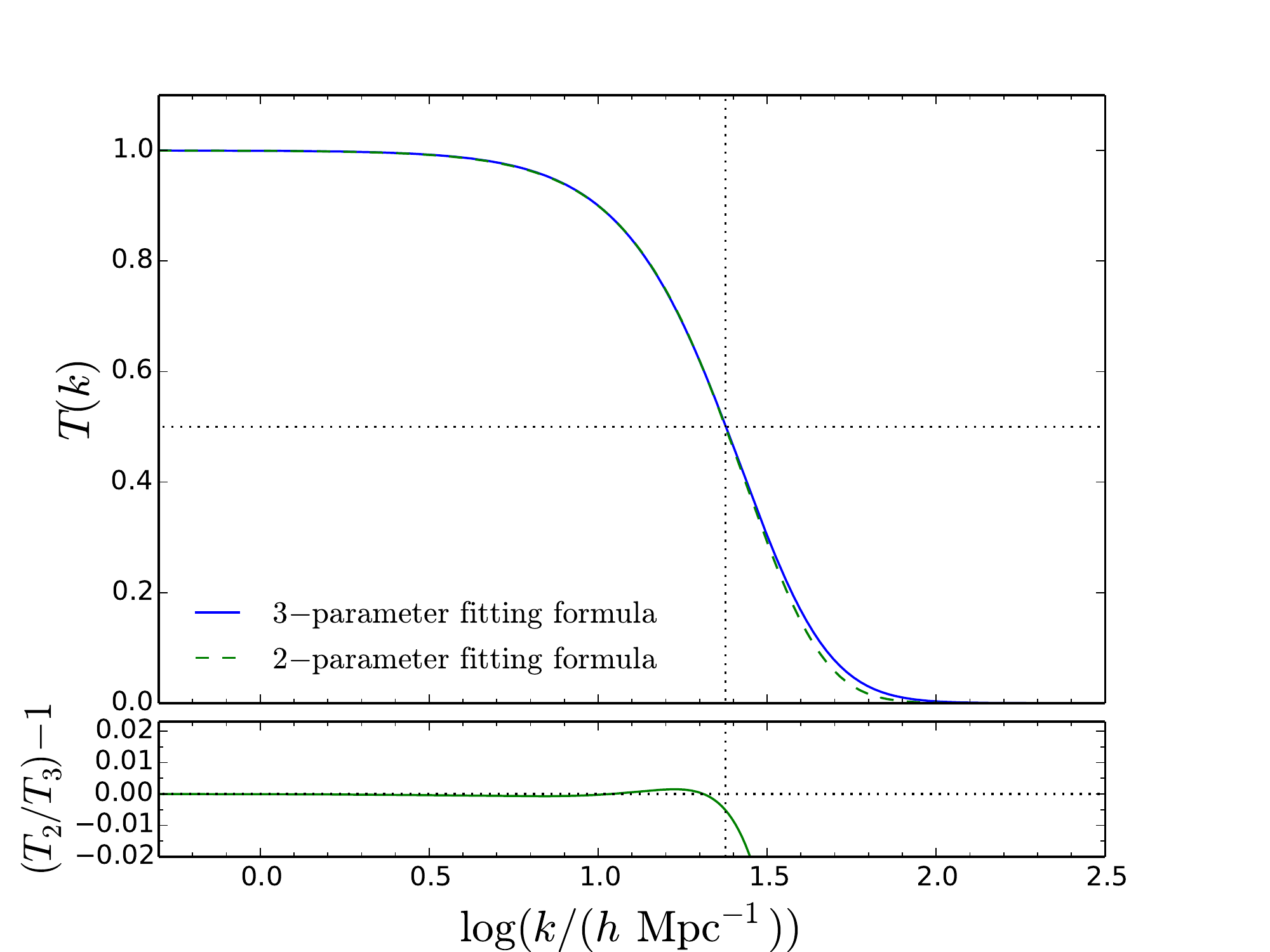}}\hspace{-2.1\baselineskip}
\subfigure[][Resonantly produced (I), power spectra.]
{\includegraphics[width=.6\textwidth]{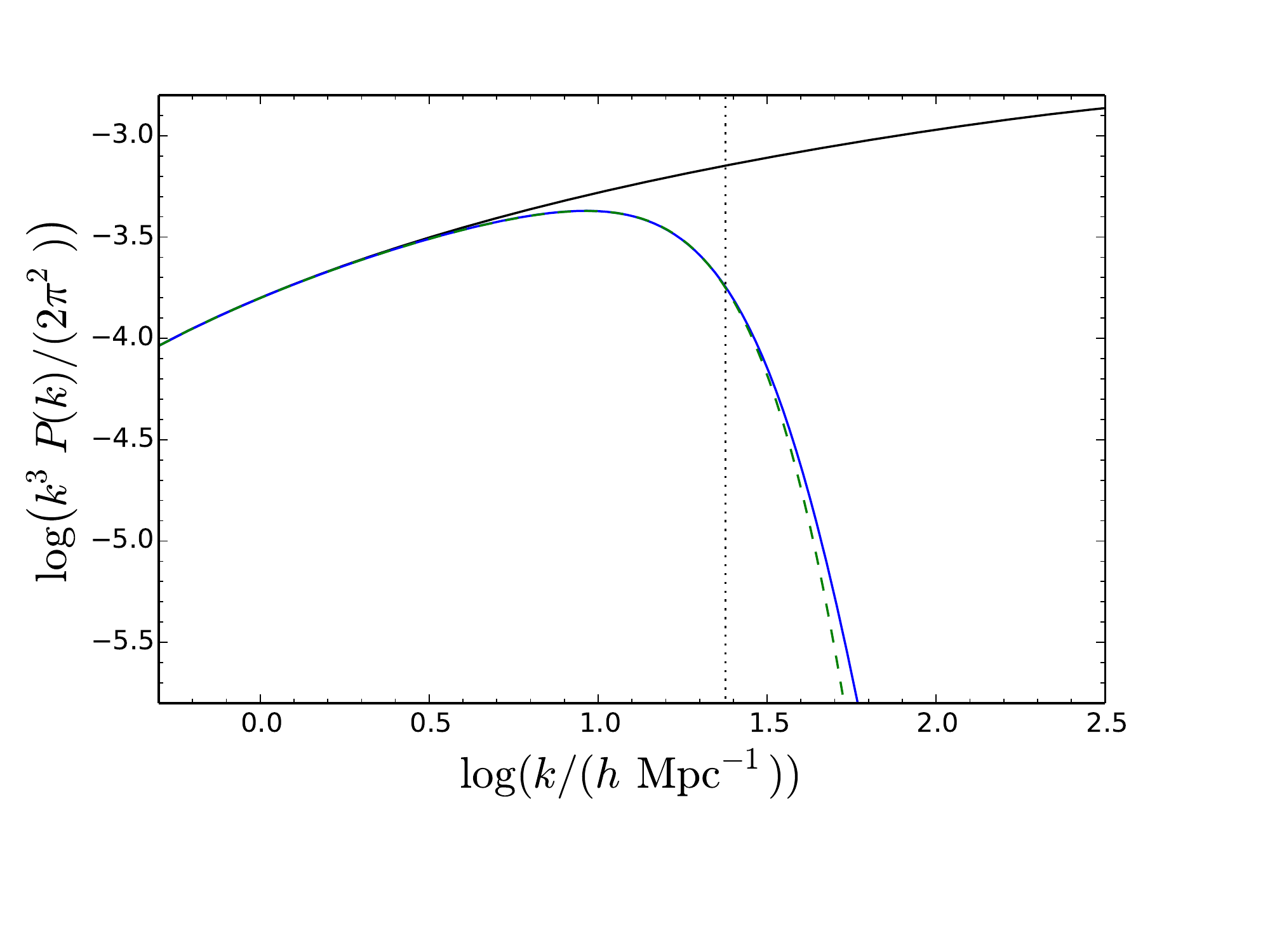}}\\
\subfigure[][Resonantly produced (II), transfer functions.]
{\includegraphics[width=.6\textwidth]{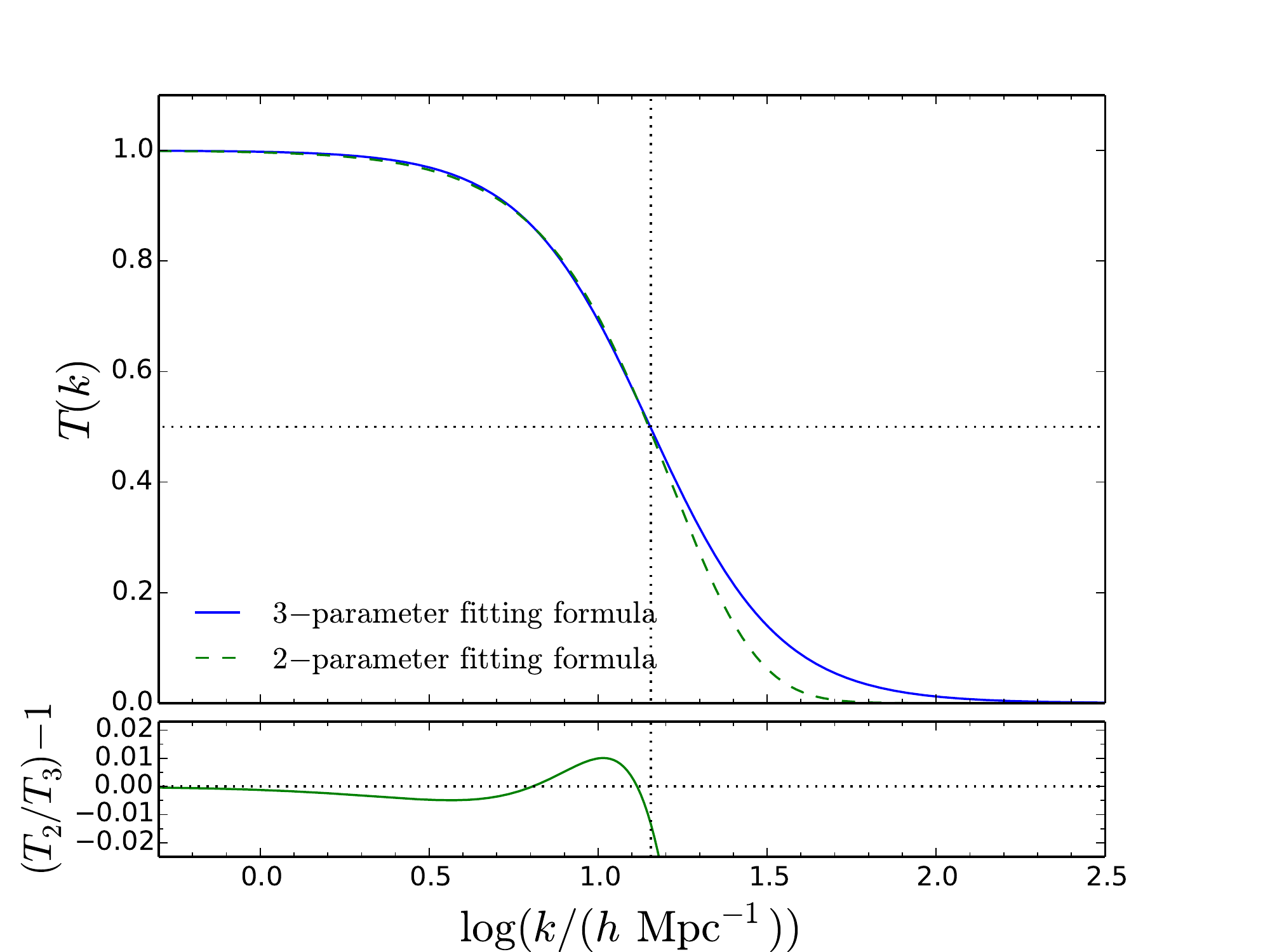}}\hspace{-2.1\baselineskip}
\subfigure[][Resonantly produced (II), power spectra.]
{\includegraphics[width=.6\textwidth]{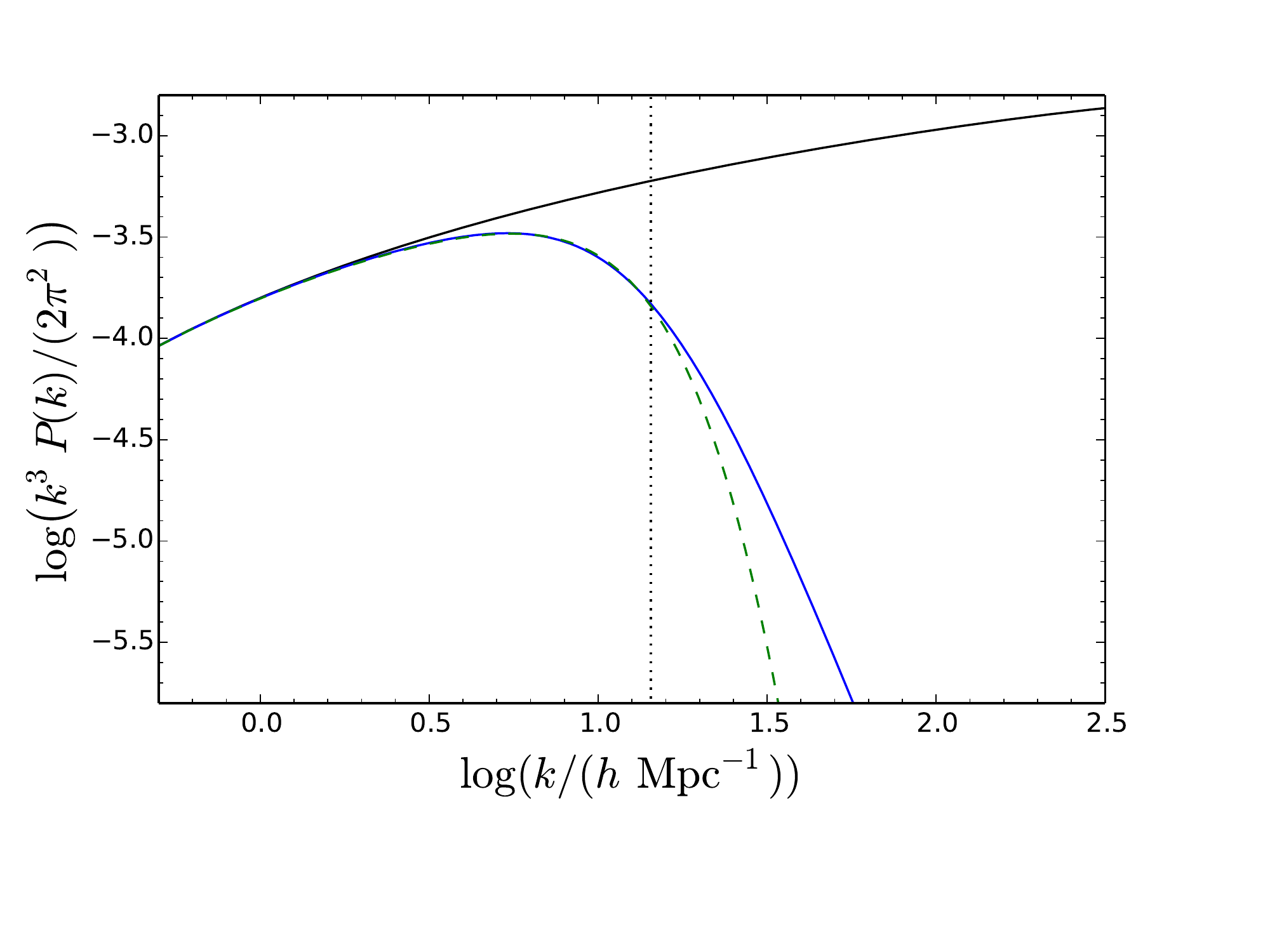}}\\
\caption{Transfer functions (left) and power spectra (right) generated using the values in Table~\ref{tab:param} for the 3-parameter (solid blue) and 2-parameter (dashed green) transfer function. The bottom panels of the figures on the left show the relative differences between the two parametrisations.  The vertical dotted line indicates the half-mode wavenumber $k_{1/2}$.}
\label{fig:fit231}
\end{figure}

\begin{figure}[]
\advance\leftskip-1.cm
\advance\rightskip-2cm
\subfigure[][Scalar decay, transfer functions]
{\includegraphics[width=.6\textwidth]{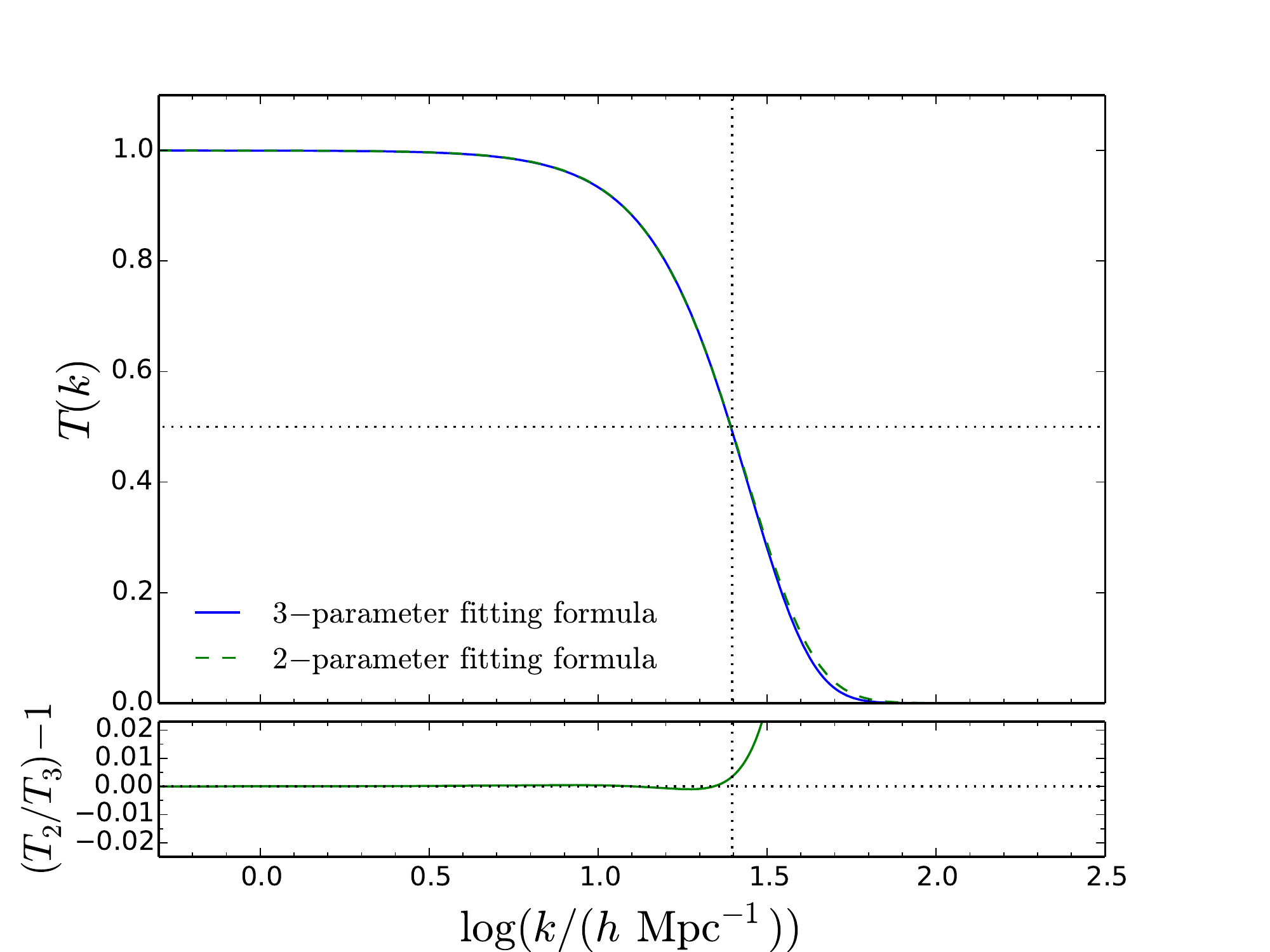}}\hspace{-2.1\baselineskip}
\subfigure[][Scalar decay, power spectra]
{\includegraphics[width=.6\textwidth]{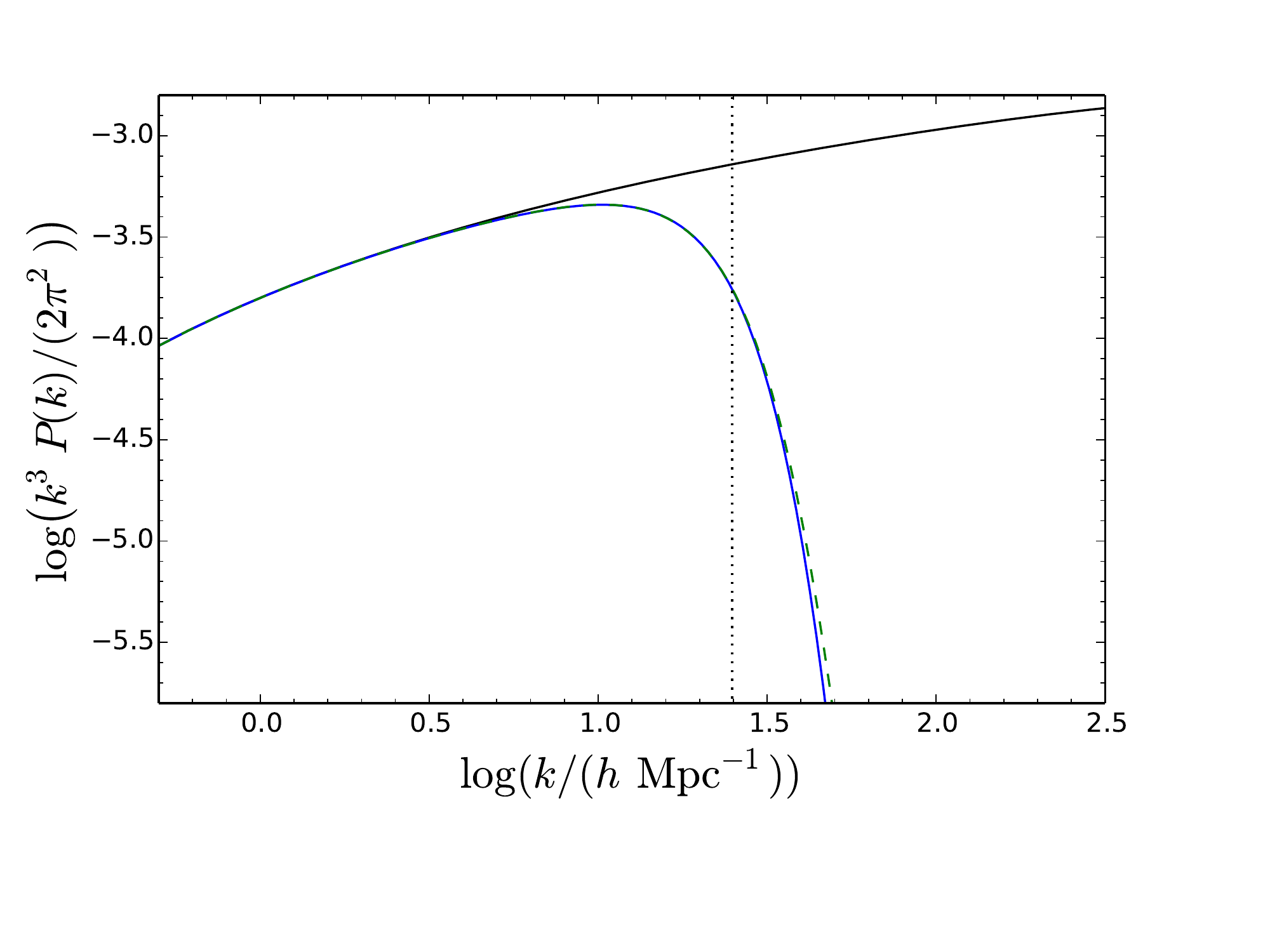}}\\
\subfigure[][Non-resonantly produced, transfer functions.]
{\includegraphics[width=.6\textwidth]{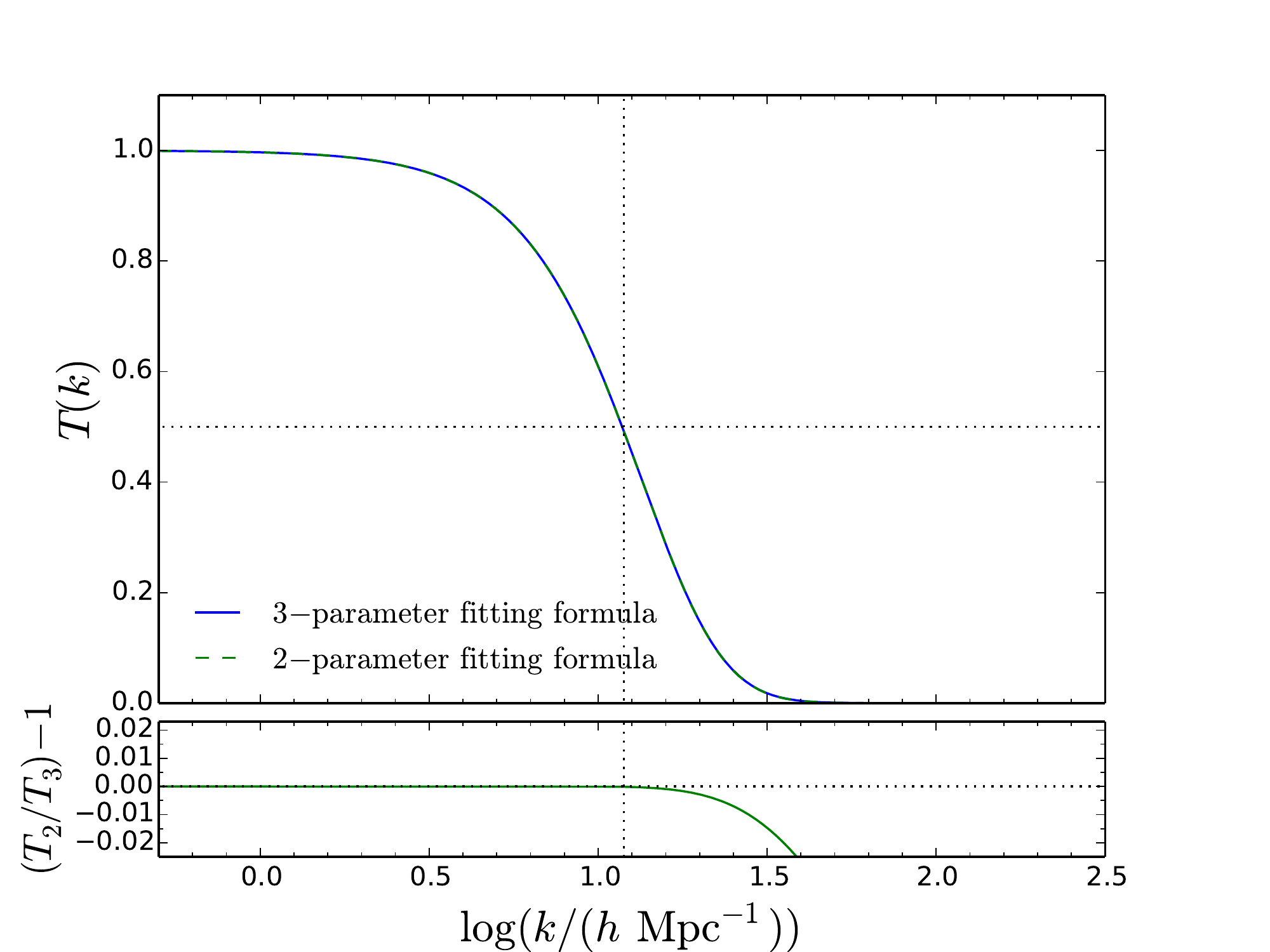}}\hspace{-2.1\baselineskip}
\subfigure[][Non-resonantly produced, power spectra.]
{\includegraphics[width=.6\textwidth]{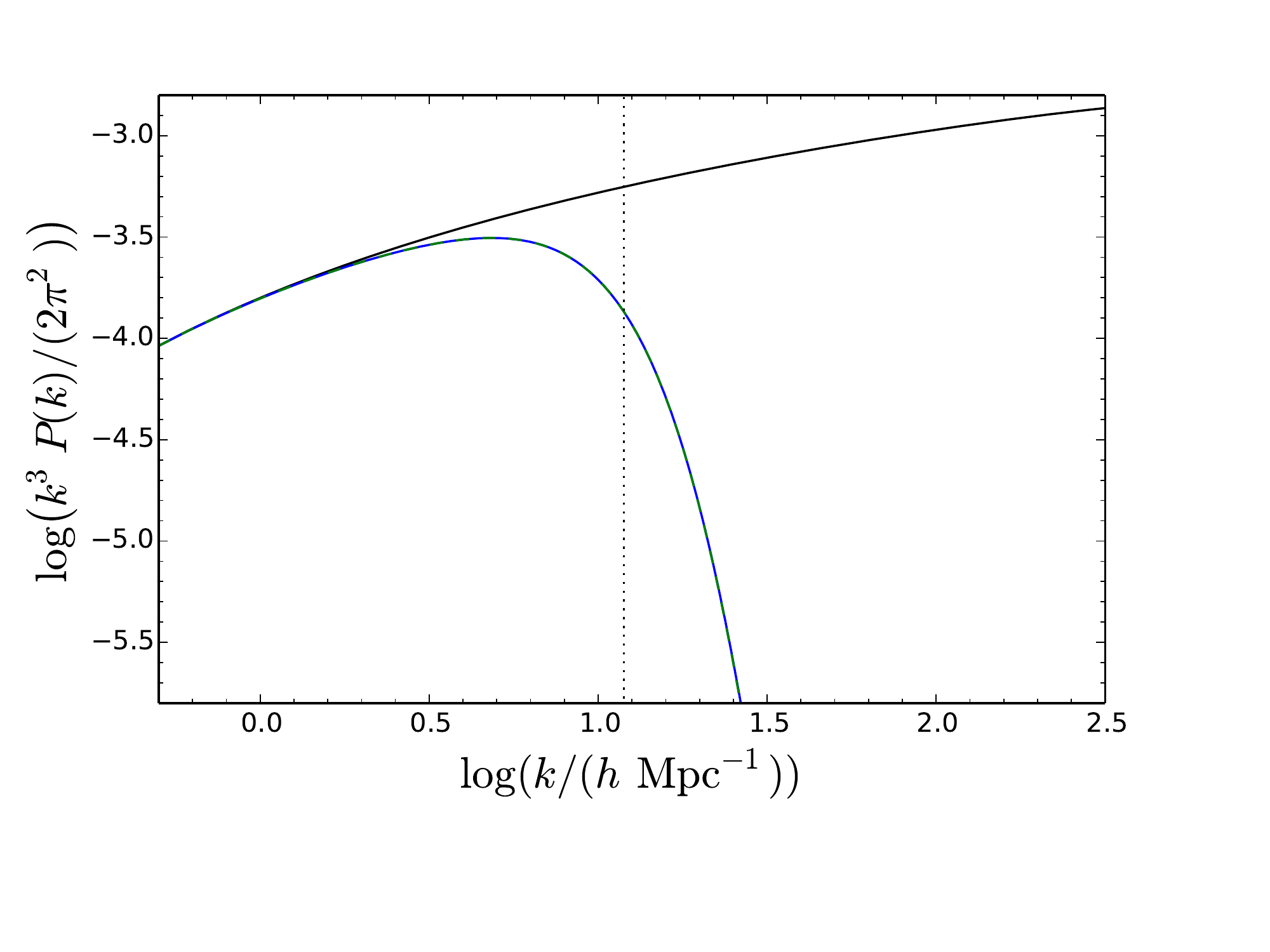}}\\
\caption{Transfer functions (left) and power spectra (right) generated using the values in Table~\ref{tab:param} for the 3-parameter (solid blue) and 2-parameter (dashed green) transfer function. The bottom panels of the figures on the left show the relative differences between the two parametrisations. The vertical dotted line represents the position of the half-mode wavenumber $k_{1/2}$.}
\label{fig:fit232}
\end{figure}

We fit the 2-parameter transfer function against some of the 3-parameter transfer functions presented in \cite{Murgia:2017lwo} using a least-squares approach and requiring  that the best-fit is obtained for $k\leq k_{1/2}$, while it does not matter if the two parametrisations diverge at higher wavenumbers. In Table~\ref{tab:param}, we show how the values of the parameters found in  \cite{Murgia:2017lwo} for a 3-parameter transfer function change when using our 2-parameter function. The plots of the transfer functions for the models in Table~\ref{tab:param} are shown in Figures~\ref{fig:fit231} and \ref{fig:fit232}. In these plots, we show the transfer functions on the left and the corresponding linear power spectra on the right. As shown in these figures, our parametrisation matches very well the parametrisation in \cite{Murgia:2017lwo} for $k\leq k_{1/2}$, and only at high wavenumbers do the two formulae diverge. Indeed, looking at the relative differences $T_2/T_3 -1$ (bottom panels in Figures~\ref{fig:fit231} and \ref{fig:fit232}), where $T_2$ and $T_3$ refer to the 2- and 3-parameter transfer functions respectively, for $k\leq k_{1/2}$ the transfer functions agree to better than $1\%$.  

\begin{figure}[h]
\centering

{\includegraphics[width=.8\textwidth]{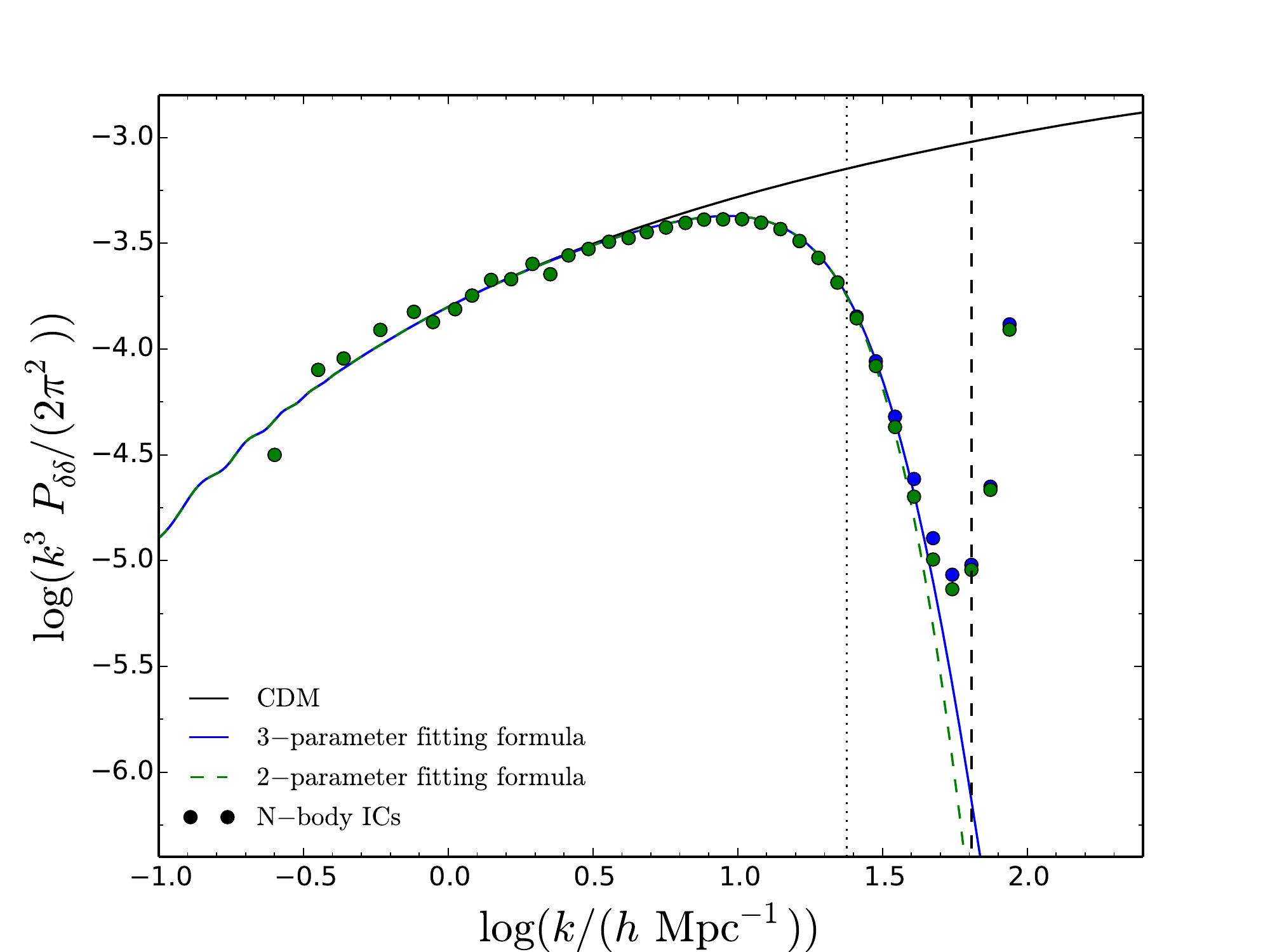}}
\caption{Initial linear matter power spectra generated at $z=199$ for resonantly produced (I) (see Table~\ref{tab:param}) using 3-parameter (blue) and 2-parameter (green) transfer function. The symbols represent the matter power spectra measured from the ICs. The black vertical dotted line represents the half-mode wavenumber $k_{1/2}$, while the black vertical dashed line is the Nyquist frequency of the simulation.}
\label{fig:ICspar}
\end{figure}

\begin{figure}[h]
\centering
\subfigure[][]
{\includegraphics[width=.6\textwidth]{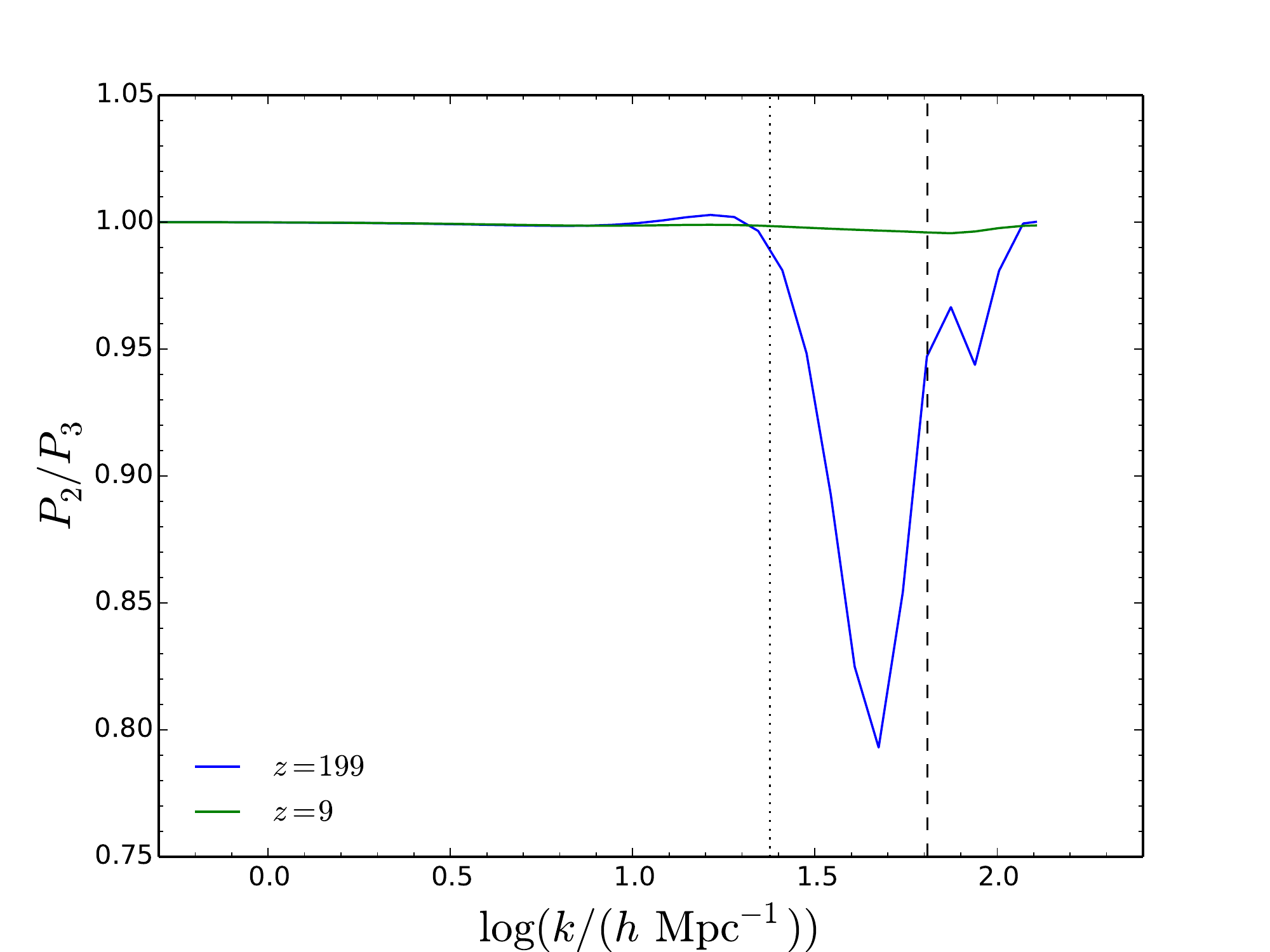}\label{fig:evolpara}}
\subfigure[][]
{\includegraphics[width=.6\textwidth]{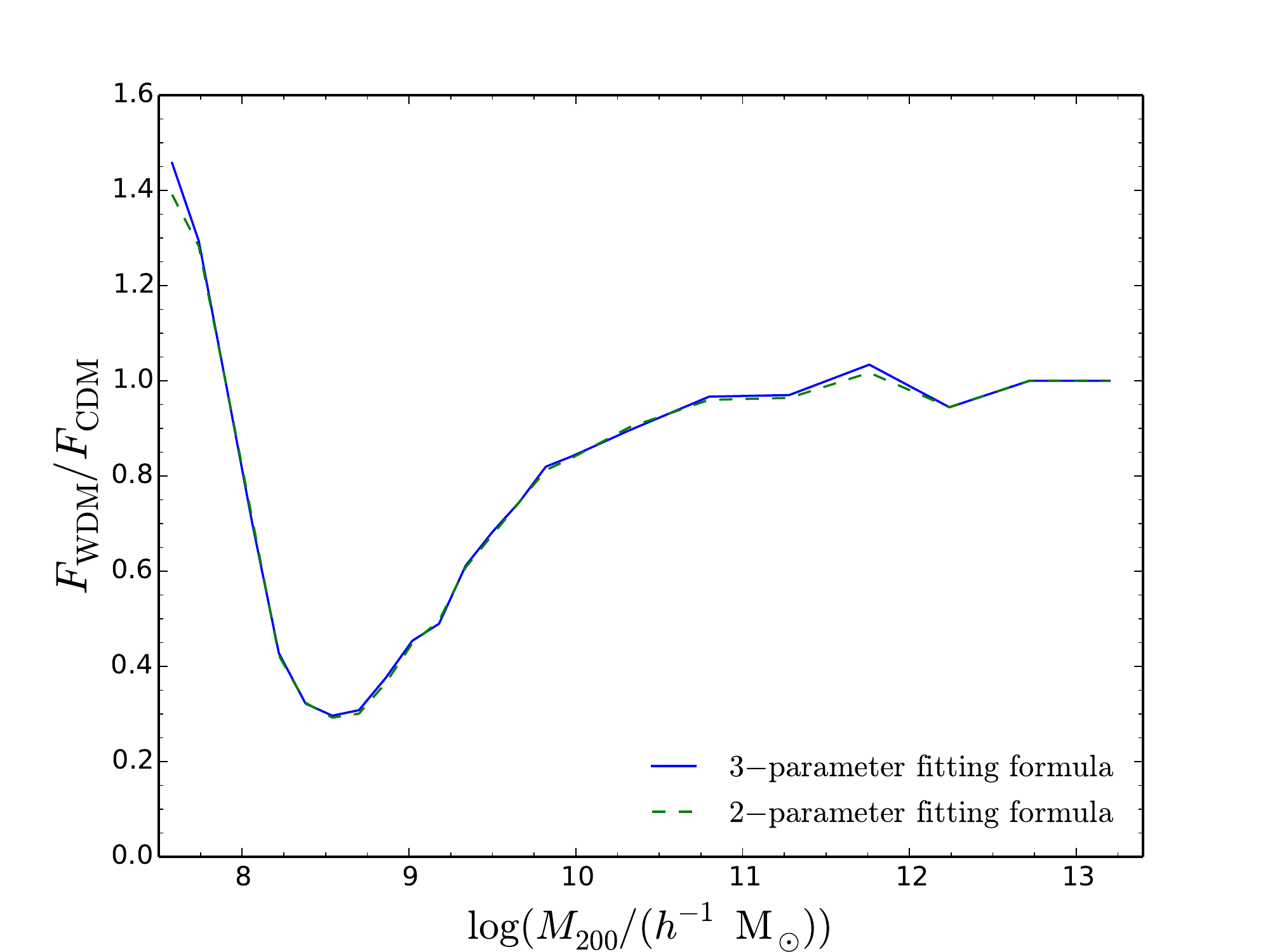}\label{fig:evolparb}}
\caption{(a) Ratios between the matter power spectra coming from the 2- and 3-parameter transfer function (as labelled) measured from N-body simulations at redshift $z=199$ and $9$. (b) Ratios of the halo mass function measured from N-body simulations at $z=0$ for both of the parametrisations (as labelled) with respect to that in CDM.}
\end{figure}

To  confirm that our parametrisation is sufficiently accurate to study DM models from the point of view of structure formation, we choose one of the above examples (the one called ``resonantly produced (I)'' in Table~\ref{tab:param}) and use N-body simulations to evolve the ICs generated at $z=199$ generated using both the parametrisations. The simulations are performed in a cubic box of length $L=25 h^{-1}$ Mpc using $N=512^3$ particles. The matter power spectra measured from the ICs are shown in Figure~\ref{fig:ICspar}, which captures the small differences between the two parametrisations at $k>k_{1/2}$. However, when the system evolves, these differences are reduced and become negligible at late times. Indeed, in Figure~\ref{fig:evolpara}, where we display the ratio $P_2/P_3$ between the evolved power spectra obtained from 2- and 3-parameter transfer functions, we see that e.g. at $z=9$ the differences are washed out and the two power spectra are identical. This is true also for lower redshifts which are not shown here.
In Figure~\ref{fig:evolparb}, we show the ratio of the halo mass functions at $z=0$ measured from the two simulations with respect to that from the CDM simulation. In this case also there are no appreciable differences between the two parametrisations. 

In conclusion, reducing the number of free parameters required to describe the damped linear theory power spectra by neglecting the high wavenumber behaviour of the transfer function does not introduce any appreciable deviations in the nonlinear matter power spectrum and the halo mass function with respect to the results coming from the full 3-parameter transfer function (at least for linear $P(k)$ with $k_{1/2}$ similar or above those considered here). This means that our parametrisation is able to capture the interesting features of a linear matter power spectrum from the point of view of structure formation.

\section{Conclusions}
Different alternatives to standard $\Lambda$CDM have been proposed to improve the performance of the model on small scales. Some of these arise from a modification of the primordial power spectrum, while others involve non-standard DM mechanisms. All such models are characterised by a damping of the matter fluctuations at high wavenumbers so that the matter power spectrum displays a cut-off on small scales. The slope of the linear power spectrum and the position of the cut-off strictly depend on the particular model considered. However, nonlinear evolution of structure transfers power from large to small scales, reducing the differences between different damped models at later times and power spectra with different slopes can yield the same cosmological structure.

We have investigated how much information is retained at late times from the initial linear power spectrum following the nonlinear growth of structure. To do this we have run a series of N-body simulations considering initial linear matter power spectra with different shapes.  We found that at late times when the system has  undergone nonlinear evolution, the shape of the initial linear theory power spectrum above the half-mode wavenumber $k_{1/2}$ is irrelevant for determining the form of the nonlinear power spectrum. Two models, whose linear power spectra are identical at small wavenumbers and differ only for $k>k_{1/2}$, will produce identical evolved power spectra at late times. On the other hand, some differences can be still seen in the halo mass function even at $z=0$. We found that this quantity is more sensitive to the linear matter power spectrum, so  potentially it can be used to  detect variations in the linear theory power at wavenumbers $k\gtrsim k_{1/2}$. However, if two linear power spectra are very similar  to one another around $k_{1/2}$ (so no big jumps near or above $k_{1/2}$), the deviations in the halo mass functions for such models are negligible. We stress that all these results are strictly valid for damped models with half-mode wavenumbers similar or above that of a linear $P(k)$ for a thermal WDM candidate with mass $m_\mathrm{WDM}=2$ keV. However, models with linear power spectra displaying cut-offs at smaller wavenumbers than those considered here are strongly excluded by Lyman-$\alpha$ constraints \cite{Viel:2013apy,Irsic:2017ixq}.

Motivated by our results, we have reduced the number of free parameters in the 3-parameter analytic fitting formula given in \cite{Murgia:2017lwo} to parametrise a damped linear power spectrum. Indeed, we have shown that a 2-parameter transfer function (which matches extremely accurately the 3-parameter transfer function at $k\leq k_{1/2}$, but which gives rise to different linear spectra at higher wavenumbers) is capable of capturing the main features of a damped model in structure formation. In particular, the halo mass function (which is more sensitive to variations in the linear theory power) seems to be unaffected by this change in the parametrisation. Indeed, the differences in the halo mass function measured from simulations of 2- and 3-parameter models are small compared with the sensitivity expected for future observations to discriminate between two damped models, such as strong lensing (see e.g. \cite{Li:2015xpc}). So, these observations cannot distinguish between the results from 2- and 3- parameter models.

Damped models come from very different underlying physical models, but if two models are characterised by linear power spectra which are very similar below the half-mode wavenumber, the results in terms of structure formations are similar. This limits what we can hope to learn about the mechanisms that occurred in the early Universe by measuring cosmic large-scale structure. Nevertheless, this simplifies the work of finding constraints on the impact of damped models of structure formation, because results coming from a particular model can be easily generalised to other models with similar linear power spectra at small wavenumbers.

As a final comment, halo statistics for damped models can be inferred using analytical techniques, such as the Press-Schechter (PS). The PS approach for damped models is characterised by a sharp-$k$ space filter (instead of a real-space spherical top-hat filter used for standard $\Lambda$CDM), whose free parameter is chosen by fitting against simulations (see e.g. \cite{Schneider:2013ria,Schneider:2014rda}). However, we have not considered this method here, because we have found that, when applying the PS approach with a sharp-$k$ filter to the damped models presented here, it does not accurately follow the behaviour of the halo mass function at small halo mass scales (when compared with N-body results). We refer to a companion study \cite{Leo:2018odn}, where we have explored in detail the results from the PS approach in the case of damped models and where we have provided a solution to the above-mentioned problem at small halo masses by introducing a new filter function.

\section*{Acknowledgements}
We thank Jianhua He for valuable discussions about power spectrum measurement. ML and SP are supported by the European Research Council under ERC Grant ``NuMass'' (FP7- IDEAS-ERC ERC-CG 617143). BL is supported by an European Research Council Starting Grant (ERC-StG-716532-PUNCA). CMB and BL acknowledge the support of the UK STFC Consolidated Grants (ST/P000541/1 and ST/L00075X/1) and Durham University. SP acknowledges partial support from the Wolfson Foundation and the Royal Society and also thanks SISSA  and IFT UAM-CSIC for support and hospitality during part of this work. SP, CMB and BL are also supported in part by the European Union's Horizon 2020 research and innovation program under the Marie Sk\l{}odowska-Curie grant agreements No. 690575 (RISE InvisiblesPlus) and 674896 (ITN Elusives). This work used the DiRAC Data Centric system at Durham University, operated by the Institute for Computational Cosmology on behalf of the STFC DiRAC HPC Facility (\href{www.dirac.ac.uk}{www.dirac.ac.uk}). This equipment was funded by BIS National E-infrastructure capital grant ST/K00042X/1, STFC capital grants ST/H008519/1 and ST/K00087X/1, STFC DiRAC Operations grant  ST/K003267/1 and Durham University. DiRAC is part of the National E-Infrastructure.



\begin{thebibliography}{99}
\bibitem{Weinberg:2013aya} D. H. Weinberg, J. S. Bullock, F. Governato, R. K. de Naray, A. H. G. Peter, \emph{Cold dark matter: controversies on small scales}, Proc. Nat. Acad. Sci. {\bf 112}, 12249-12255?? (2014) [\href{https://arxiv.org/abs/1306.0913}{arXiv:1306.0913} [astro-ph.CO]].
\bibitem{Dubinski:1991bm} J. Dubinski, R. G. Carlberg, \emph{The Structure of cold dark matter halos}, Astrophys. J. {\bf 378}, 496 (1991).
\bibitem{Navarro:1995iw} J. F. Navarro, C. S Frenk, S. D. M. White, \emph{The Structure of cold dark matter halos}, Astrophys. J. {\bf 462}, 563-575 (1996) [\href{https://arxiv.org/abs/astro-ph/9508025}{arXiv:astro-ph/9508025}]. 
\bibitem{Navarro:1996gj} J. F. Navarro, C. S Frenk, S. D. M. White, \emph{A Universal density profile from hierarchical clustering}, Astrophys. J. {\bf 490}, 493-508 (1997)  [\href{https://arxiv.org/abs/astro-ph/9611107}{arXiv:astro-ph/9611107}]. 
\bibitem{deBlok:2001hbg} W. J. G. de Blok,  S. S. McGaugh,  A. Bosma, V. C. Rubin, \emph{Mass density profiles of LSB galaxies}, Astrophys. J. {\bf 552}, L23-L26 (2001)  [\href{https://arxiv.org/abs/astro-ph/0103102}{arXiv:astro-ph/0103102}].
\bibitem{Salucci:2011ee} P. Salucci, M. I. Wilkinson, M. G. Walker, G. F. Gilmore, E. K. Grebel, A. Koch,  C. Frigerio Martins, R. F. G. Wyse, \emph{Dwarf spheroidal galaxy kinematics and spiral galaxy scaling laws}, Mon. Not. Roy. Astron. Soc. {\bf 420}, 2034 (2012)  [\href{https://arxiv.org/abs/1111.1165}{arXiv:1111.1165} [astro-ph.CO]].
\bibitem{Klypin:1999uc}  A. Klypin, A. V. Kravtsov,  O. Valenzuela, F. Prada, \emph{Where are the missing Galactic satellites?}, Astrophys. J. {\bf 522}, 82-92 (1999)  [\href{https://arxiv.org/abs/astro-ph/9901240}{arXiv:astro-ph/9901240}]. 
\bibitem{Moore:1999nt}  B. Moore, S. Ghigna, F. Governato, G. Lake, T. Quinn, J. Stadel, P. Tozzi, \emph{Dark matter substructure within galactic halos}, Astrophys. J. {\bf 524}, L19-L22 (1999)  [\href{https://arxiv.org/abs/astro-ph/9907411}{arXiv:astro-ph/9907411}].
\bibitem{2011MNRAS.415L..40B}  M. Boylan-Kolchin, J. S. Bullock,  M. Kaplinghat, \emph{Too big to fail? The puzzling darkness of massive Milky Way subhaloes}, Mon. Not. Roy. Astron. Soc. {\bf 415}, L40 (2011)  [\href{https://arxiv.org/abs/1103.0007}{arXiv:1103.0007} [astro-ph.CO]].
\bibitem{Mashchenko:2007jp} S. Mashchenko, J. Wadsley, H. M. P. Couchman, \emph{Stellar Feedback in Dwarf Galaxy Formation}, Science {\bf 319}, 174 (2008)  [\href{https://arxiv.org/abs/0711.4803}{arXiv:0711.4803} [astro-ph]].
\bibitem{2012MNRAS.421.3464P}  A. Pontzen, F. Governato, \emph{How supernova feedback turns dark matter cusps into cores}, Mon. Not. R. Astron. Soc. {\bf 421}, 3464-3471 (2012)   [\href{https://arxiv.org/abs/1106.0499}{arXiv:1106.0499} [astro-ph.CO]].
\bibitem{2013MNRAS.432.1947M} D. Martizzi, R. Teyssier, B. Moore, \emph{Cusp-core transformations induced by AGN feedback in the progenitors of cluster galaxies}??, Mon. Not. R. Astron. Soc. {\bf 432}, 1947-1954 (2013)  [\href{https://arxiv.org/abs/1211.2648}{arXiv:1211.2648} [astro-ph.CO]].
\bibitem{2014ApJ...786...87B} A. M. Brooks, A. Zolotov, \emph{Why Baryons Matter: The Kinematics of Dwarf Spheroidal Satellites}, ApJ {\bf 786}, 87 (2014)  [\href{https://arxiv.org/abs/1207.2468}{arXiv:1207.2468} [astro-ph.CO]].
\bibitem{2012MNRAS.424.2715W} J. Wang, C. S. Frenk, J. F. Navarro, L. Gao, T. Sawala, \emph{The missing massive satellites of the Milky Way}, Mon. Not. R. Astron. Soc. {\bf 424},  2715-2721 (2012)   [\href{https://arxiv.org/abs/1203.4097}{arXiv:1203.4097} [astro-ph.GA]].
\bibitem{Kamionkowski:1999vp} M. Kamionkowski, A. R. Liddle, \emph{The Dearth of halo dwarf galaxies: Is there power on short scales?}, Phys. Rev. Lett.  {\bf 84}, 4525 (2000) [\href{https://arxiv.org/abs/astro-ph/9911103}{arXiv:astro-ph/9911103}].
\bibitem{White:2000sy} M. J. White, R. A. C. Croft, \emph{Suppressing linear power on dwarf galaxy halo scales}, Astrophys. J.  {\bf 539}, 497 (2000) [\href{https://arxiv.org/abs/astro-ph/0001247v2}{arXiv:astro-ph/0001247}].
\bibitem{Yokoyama:2000tz} J. Yokoyama, \emph{Inflation and the dwarf galaxy problem}, Phys. Rev. D {\bf 62}, 123509 (2000) [\href{https://arxiv.org/abs/astro-ph/0009127}{arXiv:astro-ph/0009127}].  
\bibitem{Zentner:2002xt} A. R. Zentner, J. S. Bullock, \emph{Inflation, cold dark matter, and the central density problem}, Phys. Rev. D {\bf 66}, 043003 (2002) [\href{https://arxiv.org/abs/astro-ph/0205216}{arXiv:astro-ph/0205216}].
\bibitem{Ashoorioon:2006wc}  A. Ashoorioon, A. Krause, \emph{Power Spectrum and Signatures for Cascade Inflation}. [\href{https://arxiv.org/abs/hep-th/0607001}{arXiv:hep-th/060700}].
\bibitem{Kobayashi:2010pz} T. Kobayashi, F. Takahashi, \emph{Running Spectral Index from Inflation with Modulations}, JCAP {\bf 1101}, 026 (2011) [\href{https://arxiv.org/abs/1011.3988}{arXiv:1011.3988} [astro-ph.CO]].
\bibitem{Nakama:2017ohe} T. Nakama, J. Chluba, M. Kamionkowski, \emph{Shedding light on the small-scale crisis with CMB spectral distortions}, Phys. Rev. D {\bf 95}, no. 12, 121302 (2017)  [\href{https://arxiv.org/abs/1703.10559v2}{arXiv:1703.10559} [astro-ph.CO]].
\bibitem{Bode:2000gq} P. Bode, J. P. Ostriker, N. Turok, \emph{Halo formation in warm dark matter models}, Astrophys. J. {\bf 556}, 93-107 (2001)  [\href{http://arxiv.org/abs/astro-ph/0010389v3}{arXiv:astro-ph/0010389}].
\bibitem{Colin:2000dn} P. Colin, V. Avila-Reese, O. Valenzuela, \emph{Substructure and halo density profiles in a warm dark matter cosmology}, Astrophys. J. {\bf 542}, 622-630 (2000)  [\href{https://arxiv.org/abs/astro-ph/0004115}{arXiv:astro-ph/0004115}].
\bibitem{Hansen:2001zv} S. H. Hansen, J. Lesgourgues, S. Pastor, J. Silk, \emph{Constraining the window on sterile neutrinos as warm dark matter}, Mon. Not. Roy. Astron. Soc. {\bf 333}, 544-546 (2002)  [\href{https://arxiv.org/abs/astro-ph/0106108}{arXiv:astro-ph/0106108}].
\bibitem{Viel:2005qj} M.Viel, J. Lesgourgues, M. G. Haehnelt, S. Matarrese, A. Riotto, \emph{Constraining warm dark matter candidates including sterile neutrinos and light gravitinos with WMAP and the Lyman-alpha forest}, Phys. Rev. D {\bf 71}, 063534 (2005)  [\href{https://arxiv.org/abs/astro-ph/0501562}{arXiv:astro-ph/0501562}].
\bibitem{Dodelson:1993je} S. Dodelson, L. M. Widrow, \emph{Sterile-neutrinos as dark matter}, Phys. Rev. Lett. {\bf 72}, 17-20 (1994)  [\href{http://xxx.lanl.gov/abs/hep-ph/9303287}{arXiv:hep-ph/9303287}].
\bibitem{Dolgov:2000ew} A. D. Dolgov, S. H. Hansen, \emph{Massive sterile neutrinos as warm dark matter}, Astropart. Phys. {\bf 16}, 339-344 (2002)   [\href{https://arxiv.org/abs/hep-ph/0009083}{arXiv:hep-ph/0009083}].
\bibitem{Asaka:2006nq} T. Asaka, M. Laine, M. Shaposhnikov, \emph{Lightest sterile neutrino abundance within the nuMSM}, JHEP {\bf 01}, 091 (2007)  [Erratum: JHEP02,028(2015)] [\href{http://xxx.lanl.gov/abs/hep-ph/0612182}{arXiv:hep-ph/0612182}].
\bibitem{Enqvist:1990ek}  K. Enqvist, K. Kainulainen, J. Maalampi, \emph{Resonant neutrino transitions and nucleosynthesis}, Phys. Lett. B {\bf 249},  531-534 (1990).
\bibitem{Shi:1998km} X. Shi, G. M. Fuller, \emph{A New dark matter candidate: Nonthermal sterile
                        neutrinos}, Phys. Rev. Lett. {\bf 82}, 2832 (1999)  [\href{http://xxx.lanl.gov/abs/astro-ph/9810076}{arXiv:astro-ph/9810076}].
\bibitem{Abazajian:2001nj} K. Abazajian, G. M. Fuller, M. Patel, \emph{Sterile neutrino hot, warm, and cold dark matter}, Phys. Rev. D {\bf 64},  023501 (2001)  [\href{http://xxx.lanl.gov/abs/astro-ph/0101524}{arXiv:astro-ph/0101524}]. 
\bibitem{Kusenko:2006rh} A. Kusenko, \emph{Sterile neutrinos, dark matter, and the pulsar
                        velocities in models with a Higgs singlet}, Phys. Rev. Lett. {\bf 97}, 241301 (2006)  [\href{https://arxiv.org/abs/hep-ph/0609081}{arXiv:hep-ph/0609081}].
\bibitem{Petraki:2007gq} K. Petraki, A. Kusenko, \emph{Dark-matter sterile neutrinos in models with a gauge
                        singlet in the Higgs sector}, Phys. Rev. D {\bf 77}, 065014 (2008)  [\href{https://arxiv.org/abs/0711.4646}{arXiv:0711.4646} [hep-ph]].
\bibitem{Merle:2015oja} A. Merle, M. Totzauer, \emph{keV Sterile Neutrino Dark Matter from Singlet Scalar
                        Decays: Basic Concepts and Subtle Features}, JCAP {\bf 1506}, 011 (2015)  [\href{https://arxiv.org/abs/1502.01011}{arXiv:1502.01011} [hep-ph]].
\bibitem{Konig:2016dzg}  J. K\"{o}nig, A. Merle, M. Totzauer, \emph{keV Sterile Neutrino Dark Matter from Singlet Scalar
                        Decays: The Most General Case}, JCAP {\bf 1611}, 038 (2016)  [\href{https://arxiv.org/abs/1609.01289}{arXiv:1609.01289} [hep-ph]].
\bibitem{Boehm:2004th} C. Boehm, R. Schaeffer, \emph{Constraints on dark matter interactions from structure formation: Damping lengths}, Astron. Astrophys. {\bf 438}, 419-442 (2005)  [\href{https://arxiv.org/abs/astro-ph/0410591}{arXiv:astro-ph/0410591}]. 
\bibitem{Boehm:2014vja} C. Boehm, J. A. Schewtschenko, R. J. Wilkinson, C. M. Baugh, S. Pascoli, \emph{Using the Milky Way satellites to study interactions
                        between cold dark matter and radiation}, Mon. Not. Roy. Astron. Soc. {\bf 445},  L31-L35 (2014)  [\href{https://arxiv.org/abs/1404.7012}{arXiv:1404.7012} [astro-ph.CO]].
\bibitem{Schewtschenko:2014fca} J. A. Schewtschenko, R. J. Wilkinson, C. M. Baugh, C. Boehm, S. Pascoli, \emph{Dark matter-radiation interactions: the impact on dark matter haloes}, Mon.\ Not.\ Roy.\ Astron.\ Soc.\  {\bf 449}, 3587 (2015)
   [\href{https://arxiv.org/abs/1412.4905}{arXiv:1412.4905}].               
\bibitem{Spergel:1999mh} D. N. Spergel, P. J. Steinhardt, \emph{Observational evidence for selfinteracting cold dark
                        matter}, Phys. Rev. Lett. {\bf 84}, 3760-3763 (2000)  [\href{https://arxiv.org/abs/astro-ph/9909386}{arXiv:astro-ph/9909386}].
\bibitem{Marsh:2015xka} D. J. E. Marsh, \emph{Axion Cosmology}, Phys. Rept. {\bf 643}, 1-79?? (2016)  [\href{https://arxiv.org/abs/1510.07633}{arXiv:1510.07633} [astro-ph.CO]].
\bibitem{Murgia:2017lwo} R. Murgia, A. Merle, M. Viel, M. Totzauer, A. Schneider, \emph{"Non-cold" dark matter at small scales: a general approach}, JCAP {\bf 1711}, 046 (2017) [\href{https://arxiv.org/abs/1704.07838}{arXiv:1704.07838} [astro-ph.CO]].
\bibitem{2012MNRAS.421...50V} M. Viel, K. Markovic, M. Baldi, J. Weller, \emph{The Non-Linear Matter Power Spectrum in Warm Dark Matter Cosmologies}, Mon. Not. R. Astron. Soc. {\bf 421},  50-62 (2012)  [\href{https://arxiv.org/abs/1107.4094}{arXiv:1107.4094} [astro-ph.CO]].
\bibitem{Leo:2017zff} M. Leo, C. M. Baugh, B. Li, S. Pascoli, \emph{The Effect of Thermal Velocities on Structure Formation in N-body Simulations of Warm Dark Matter}, JCAP {\bf 11}, 017 (2017) [\href{https://arxiv.org/abs/1706.07837}{arXiv:1706.07837} [astro-ph.CO]].
\bibitem{2011arXiv1104.2932L} J. Lesgourgues, \emph{The Cosmic Linear Anisotropy Solving System (CLASS) I: Overview}, [\href{http://arxiv.org/abs/1104.2932}{arXiv:1104.2932} [astro-ph.IM]].
\bibitem{2011JCAP...09..032L} J. Lesgourgues, T. Tram, \emph{The Cosmic Linear Anisotropy Solving System (CLASS) IV: efficient implementation of non-cold relics}, JCAP {\bf 09}, 032 (2011) [\href{https://arxiv.org/abs/1104.2935}{arXiv:1104.2935}].
\bibitem{Bose:2015mga} S. Bose, W. A. Hellwing, C. S. Frenk, A. Jenkins, M. R. Lovell, J. C. Helly, B. Li, \emph{The COpernicus COmplexio: Statistical Properties of Warm Dark Matter Haloes},  Mon. Not. Roy. Astron. Soc.  {\bf 455}, 318 (2016)
   [\href{https://arxiv.org/abs/1507.01998}{arXiv:1507.01998} [astro-ph.CO]].
\bibitem{Crocce:2006ve} M. Crocce, S. Pueblas, R. Scoccimarro, \emph{Transients from Initial Conditions in Cosmological Simulations}, Mon. Not. Roy. Astron. Soc.  {\bf 373}, 369 (2006) [\href{https://arxiv.org/abs/astro-ph/0606505}{arXiv:astro-ph/0606505}].
\bibitem{Springel:2005mi} V. Springel, \emph{The cosmological simulation code GADGET-2}, Mon. Not. Roy. Astron. Soc.  {\bf 364}, 1105 (2005)
   [\href{http://arxiv.org/abs/astro-ph/0505010}{arXiv:astro-ph/0505010}].
\bibitem{2012MNRAS.420.2318L} M. Lovell, V. Eke, C. Frenk, L. Gao, A. Jenkins, T. Theuns, J. Wang, S. White, A. Boyarsky, O. Ruchayskiy, \emph{The haloes of bright satellite galaxies in a warm dark matter universe}, Mon. Not. Roy. Astr. Soc. {\bf 420}, 2318-2324 (2012) [\href{https://arxiv.org/abs/1104.2929v2}{arXiv:1104.2929} [astro-ph.CO]].
\bibitem{Schneider:2013ria}  A. Schneider, R. E. Smith, D. Reed, \emph{Halo Mass Function and the Free Streaming Scale},
  Mon. Not. Roy. Astron. Soc.  {\bf 433}, 1573 (2013) [\href{https://arxiv.org/abs/1303.0839}{arXiv:1303.0839} [astro-ph.CO]]. 
\bibitem{2013MNRAS.428..882M} A. V. Maccio', O. Ruchayskiy, A. Boyarsky, J. C. Munoz-Cuartas, \emph{The inner structure of haloes in Cold+Warm dark matter models}, Mon. Not. Roy. Astron. Soc. {\bf 428}, 882-890 (2013)  [\href{https://arxiv.org/abs/1202.2858v3}{arXiv:1202.2858} [astro-ph.CO]].
\bibitem{2013MNRAS.430.2346S} S. Shao, L. Gao, T. Theuns, C. S. Frenk, \emph{The phase-space density of fermionic dark matter haloes}, Mon. Not. R. Astron. Soc. {\bf 430}, 2346-2357 (2013)  [\href{https://arxiv.org/abs/1209.5563v2}{arXiv:1209.5563} [astro-ph.CO]].
\bibitem{2013ApJ...762..109B} P. S. Behroozi, R. H. Wechsler, H. Wu , \emph{The Rockstar Phase-Space Temporal Halo Finder and the Velocity Offsets of Cluster Cores}, ApJ {\bf 762}, 109 (2013)  [\href{https://arxiv.org/abs/1110.4372}{arXiv:1110.4372} [astro-ph.CO]].
\bibitem{Wang:2007he} J. Wang and S. D. M. White, \emph{Discreteness effects in simulations of Hot/Warm dark matter}, Mon. Not. Roy. Astron. Soc.  {\bf 380}, 93 (2007)
   [\href{https://arxiv.org/abs/astro-ph/0702575}{arXiv:astro-ph/0702575}].
\bibitem{2012MNRAS.424..684S} A. Schneider, R. E. Smith, A. V. Maccio, B. Moore, \emph{Non-linear evolution of cosmological structures in warm dark matter models}, Mon. Not. R. Astron. Soc. {\bf 424},  684-698 (2012)  [\href{https://arxiv.org/abs/1112.0330}{arXiv:1112.0330} [astro-ph.CO]].
\bibitem{Lovell:2013ola} M. R. Lovell,  C. S. Frenk, V. R. Eke, A. Jenkins, L. Gao, T. Theuns, \emph{The properties of warm dark matter haloes}, Mon. Not. Roy. Astron. Soc.  {\bf 439}, 300 (2014)
   [\href{https://arxiv.org/abs/1308.1399}{arXiv:1308.1399} [astro-ph.CO]].
\bibitem{Power:2013rpw} C. Power, 
  \emph{Seeking Observable Imprints of Small-Scale Structure on the Properties of Dark Matter Haloes},
   Publ. Astron. Soc. Austral.  {\bf 30}, 53 (2013)
   [\href{https://arxiv.org/abs/1309.1591v1}{arXiv:1309.1591} [astro-ph.CO]].
\bibitem{Schneider:2014rda} 
  A. Schneider,
  \emph{Structure formation with suppressed small-scale perturbations},
  Mon. Not. Roy. Astron. Soc.  {\bf 451}, no. 3, 3117 (2015)
  [\href{https://arxiv.org/abs/1412.2133}{arXiv:1412.2133} [astro-ph.CO]].  
\bibitem{Power:2016usj} C. Power, A. S. G. Robotham, D. Obreschkow, A. Hobbs, G. F. Lewis, 
  \emph{Spurious haloes and discreteness-driven relaxation in cosmological simulations},
   Mon. Not. Roy. Astron. Soc.  {\bf 462}, 474 (2016)
   [\href{https://arxiv.org/abs/1606.02038}{arXiv:1606.02038} [astro-ph.CO]].

\bibitem{Viel:2013apy} M. Viel, G. D. Becker, J. S. Bolton, M. G. Haehnelt, \emph{Warm dark matter as a solution to the small scale
                        crisis: New constraints from high redshift Lyman-$\alpha$ forest
                        data}, Phys. Rev. D {\bf 88}, 043502 (2013)  [\href{http://arxiv.org/abs/1306.2314v2}{arXiv:1306.2314} [astro-ph.CO]].
\bibitem{Irsic:2017ixq} V. Irsic {\it et al.}, \emph{New Constraints on the free-streaming of warm dark matter from intermediate and small scale Lyman-$\alpha$ forest data}, [\href{https://arxiv.org/abs/1702.01764}{arXiv:1702.01764} [astro-ph.CO]].
\bibitem{Li:2015xpc} 
  R. Li, C. S. Frenk, S. Cole, L. Gao, S. Bose, W. A. Hellwing,
  \emph{Constraints on the identity of the dark matter from strong gravitational lenses},
  Mon. Not. Roy. Astron. Soc. {\bf 460}, 363 (2016)
  [\href{https://arxiv.org/abs/1512.06507}{arXiv:1512.06507} [astro-ph.CO]].
\bibitem{Leo:2018odn} M. Leo, C. M. Baugh, B. Li, S. Pascoli, \emph{A new smooth-$k$ space filter approach to calculate halo abundances}, JCAP {\bf 04}, 010 (2018) [\href{https://arxiv.org/abs/1801.02547}{arXiv:1801.02547} [astro-ph.CO]].
\end{thebibliography}
\end{document}